\documentclass[prd,superscriptaddress,amsfonts,amssymb,amsmath,showpacs,twocolumn]{revtex4-2}
\usepackage{bm}
\usepackage{amsfonts}
\usepackage{latexsym}
\usepackage[latin1]{inputenc}
\usepackage{graphicx}
\usepackage{amsmath}
\usepackage{palatino}
\usepackage{mathpazo}
\usepackage{textcomp}
\usepackage{float}
\usepackage{booktabs}
\usepackage{dcolumn}
\usepackage{hyperref}
\hypersetup{colorlinks,citecolor=blue}
\usepackage{amsmath}
\usepackage{xcolor}
\usepackage{orcidlink}
\usepackage[caption=false]{subfig}
\usepackage{commath}
\def\jnl@style{\it}
\def\aaref@jnl#1{{\jnl@style#1}}

\def\aaref@jnl#1{{\jnl@style#1}}

\def\aj{\aaref@jnl{AJ}}                   
\def\apj{\aaref@jnl{ApJ}}                 
\def\apjl{\aaref@jnl{ApJ}}                
\def\apjs{\aaref@jnl{ApJS}}               
\def\apss{\aaref@jnl{Ap\&SS}}             
\def\aap{\aaref@jnl{A\&A}}                
\def\aapr{\aaref@jnl{A\&A~Rev.}}          
\def\aaps{\aaref@jnl{A\&AS}}              
\def\mnras{\aaref@jnl{Mon.~Not.~Roy.~Astron.~Soc.}}             
\def\prd{\aaref@jnl{Phys.~Rev.~D}}        
\def\prc{\aaref@jnl{Phys.~Rev.~C}}  
\def\prl{\aaref@jnl{Phys.~Rev.~Lett.}}    
\def\qjras{\aaref@jnl{QJRAS}}             
\def\skytel{\aaref@jnl{S\&T}}             
\def\ssr{\aaref@jnl{Space~Sci.~Rev.}}     
\def\zap{\aaref@jnl{ZAp}}                 
\def\nat{\aaref@jnl{Nature}}              
\def\aplett{\aaref@jnl{Astrophys.~Lett.}} 
\def\apspr{\aaref@jnl{Astrophys.~Space~Phys.~Res.}} 
\def\physrep{\aaref@jnl{Phys.~Rep.}}      
\def\physscr{\aaref@jnl{Phys.~Scr}}       
\def\commat{\aaref@jnl{Comm.~Math.~Phys.}}              
\def\science{\aaref@jnl{Science}}               
\def\cqg{\aaref@jnl{Classical Quant.~Grav.}}            
\def\jpcs{\aaref@jnl{JPCS}}                                     
\def\ijmpd{\aaref@jnl{Int.~J.~Mod.~Phys.~D}}                    
\def\grg{\aaref@jnl{Gen.~Relat.~Gravit.}}               
\def\rpp{\aaref@jnl{Rep.~Prog.~Phys.}}          
\def\npa{\aaref@jnl{Nucl.~Phys.~A}}        
\def\lrr{\aaref@jnl{Living Rev.~Rel.}}                   
\def\jcap{\aaref@jnl{J.~Cosmology Astropart.~Phys.}}    
\def\rmp{\aaref@jnl{Rev.~Mod.~Phys.}}   
\def\epjc{\aaref@jnl{Eur.~Phys.~J.~C}}

\allowdisplaybreaks[1]

\begin{document}

\color{black}       

\title{Complete dark energy scenario in $f(Q)$ gravity}

\author{Raja Solanki\orcidlink{0000-0001-8849-7688}}
\email{rajasolanki8268@gmail.com}
\affiliation{Department of Mathematics, Birla Institute of Technology and
Science-Pilani,\\ Hyderabad Campus, Hyderabad-500078, India.}
\author{Avik De\orcidlink{0000-0001-6475-3085}}
\email{avikde@utar.edu.my}
\affiliation{Department of Mathematical and Actuarial Sciences, Universiti Tunku Abdul Rahman, Jalan Sungai Long,
43000 Cheras, Malaysia}
\author{P.K. Sahoo\orcidlink{0000-0003-2130-8832}}
\email{pksahoo@hyderabad.bits-pilani.ac.in}
\affiliation{Department of Mathematics, Birla Institute of Technology and
Science-Pilani,\\ Hyderabad Campus, Hyderabad-500078, India.}

\date{\today}
\begin{abstract}

In theoretical physics, the fundamental nature and evolution mechanism of dark energy is still an open question. In General Relativity Theory, the simplest explanation for dark energy is the cosmological constant $\Lambda$. However, the cosmological constant $\Lambda$ facing a sensitive problem called fine-tunning problem. In the present work, we follow a different approach where the gravitational sector is the responsible candidate for the evolution of dark energy instead of the matter source. The modified symmetric teleparallel gravity or $f(Q)$ gravity is a recently proposed theory of gravity where the gravitational interaction ruled by the non-metricity term $Q$. In this manuscript, we assume a $f(Q)$ model that contains a linear and a non-linear form of non-metricity scalar, particularly $f(Q)=\alpha Q + \beta Q^n$, where $\alpha$, $\beta$, and $n$ are free model parameters. Then we find the values of our model parameters that would be in agreement with the observed value of cosmographic parameters. We analyze the behavior of different cosmological parameters like deceleration parameter, density, and the equation of state parameter with the energy conditions for our cosmological model. We found that for higher positive values of $n$ specifically $n \geq 1$, dark energy fluid part evolving due to non-metricity behaves like quintessence type dark energy while for higher negative values of $n$ specifically $n \leq -1$ it follows phantom scenario. Further for $n=0$, our cosmological $f(Q)$ model behaves like $\Lambda$CDM model. Thus, we conclude that the geometrical generalization of GR can be a viable candidate for the description of origin of the dark energy.

\end{abstract}

\maketitle

\section{Introduction}\label{sec1}

Several observational results from type Ia Supernova \cite{Riess,Perlmutter}, large scale structure \cite{T.Koivisto,S.F.},  Cosmic Microwave Background \cite{R.R.,Z.Y.}, Baryonic Acoustic Oscillations \cite{D.J.,W.J.}, and the WMAP experiment \cite{C.L.,D.N.} 
indicate an accelerating behavior of the universe. This give rise to several theories in order to describe the recent observed acceleration. In general relativity theory (GR), an unknown form of energy that possesses high negative pressure called dark energy (DE) drives this accelerated expansion. The final fate of this accelerating expansion is completely depends on the fundamental nature of dark energy. The cosmological constant $\Lambda$ in GR is the most successful description of the dark energy so far and it is represented by $\omega=-1$. Although, there is a much inconsistency between the value of the cosmological constant $\Lambda$ acquired by the observations and its value obtained from quantum gravity \cite{S.W.}. This discrepancy in the value of cosmological constant is known as cosmological constant problem. Furthermore, the present density of dark energy and the density of non-relativistic matter happen to be the same order of magnitude while their evolution are different. This observed coincidence between the densities referred to as cosmic coincidence problem. Another widely analyzed  DE model is the model with quintessence dark energy which is characterized by $-1<\omega<-\frac{1}{3}$ \cite{RP,M.T.}. In quintessence scenario, dark energy density decreases with cosmic time \cite{LX}. Further, the least theoretically understood DE is the phantom energy that is characterized by $\omega<-1$. Nowadays researchers attracted towards the phantom energy case due to its unusual charateristics. In phantom scenario, dark energy density increases with time that results a finite-time future singularity. For the classification of singularities, see the references \cite{KI,DB,DB-2}. It also a cause of violation of all the four energy conditions that is necessary to constrain wormholes \cite{ZH}. To bypass the undetected dark energy, modified gravity theories were proposed in the literature as an alternative to GR. Such an approach has been widely discussed in the references \cite{L.A.,SA,R.F.}. In the present work, we follow a different approach where the gravitational sector is the responsible candidate for the evolution of dark energy instead of the matter source. We investigate different dark energy scenarios in the framework of $f(Q)$ gravity, where $Q$ is the non-metricity scalar. The modified symmetric teleparallel gravity or $f(Q)$ gravity has been recently proposed by J. B. Jim\'enez et al \cite{J.B.}.

The curvature representation of gravity is the most prominent notions in natural sciences. Apart from the curvature representation of the spacetime's geodesic structure, the torsion and the non-metricity are two geometrical objects that can successfully establish the gravity. In General Relativity, a metric-compatible and torsion-free connection namely Levi-Civita connection determines the gravity with the help of spacetime curvature. As a consequence, it convince that the geodesics are autoparallel in this case. The second notion benefits from a  curvature-free, metric-compatible connection with the presence of torsion, which is known as the Teleparallel Equivalent of GR (TEGR). Whereas the last one makes use of a curvatureless and torsionless connection with the non-metricity, and is called the Symmetric Teleparallel Equivalent of GR (STEGR). As discussed in reference \cite{Tom}, with the help of suitable choice of coordinates connection can be globally cut off from the discussion. Thus the geometrical framework in the last setting is the most simplest among all the three equivalent descriptions of gravity. This appropriate choice of coordinates is referred to as the coincident guage coordinates. Moreover, in this particular coordinates system, it entails only the first derivatives of the metric tensor, unlike the standard theory of gravity, and thus generates a well-posed variational principle without any Gibbons-Hawking-York (GHY) boundary terms. The generalization of GR give rise to $f(R)$ gravity theory in which the gravtiational interaction attributed to the non-vanishing curvature with vanishing torsion and non-metricity \cite{A.A.} whereas the generalization of the teleparallel equivalent of GR known as $f(\mathcal{T})$ gravity theory in which the space-time is described by the non-zero torsion with vanishing curvature and non-metricity \cite{GB}. The $f(Q)$ theory of gravity is a generalization of the symmetric teleparallel equivalent of GR in which  the non-metricity scalar describes the gravitational interactions with zero curvature and torsion. Due to the nature of the connection terms associated with torsion and non-metricity, one can consider non-linear Lagrangians without having the problem of obtaining higher order terms in the field equations. Ofcourse, $f(\mathcal{T})$ theory is more restrictive than $f(Q)$ as the corresponding connection has to satisfy more conditions, in the sense, more independent equations involved in $Q_{abc}=0$ than $\mathcal{T}_{abc}=0$ in $f(Q)$ theory. The presence of anti-symmetric part in the field equations of the $f(\mathcal{T})$ theory and the local Lorentz invariance problem hinder the Bianchi identity to hold automatically whereas the field equations in $f(Q)$ theory is purely symmetric and free of this issue.

Several investigations have been done in the context of $f(Q)$ gravity with different aspects such as energy conditions \cite{MS},  cosmography \cite{MS-2}, covariant formulation \cite{DZ}, spherically symmetric configuration \cite{RHL}, and signature of $f(Q)$ gravity \cite{NF}. The growth index of matter perturbations have been studied in the context of $f(Q)$ \cite{WK}. Harko et al. analyzed the coupling matter in modified $Q$ gravity \cite{HRK}. Dimakis et al. studied 
quantum cosmology for a $f(Q)$ polynomial model \cite{ND}. The geodesic deviation equation in $f(Q)$ gravity has been studied and some fundamental results were obtained in \cite{GDV}. F. K. Anagnostopoulos has presented an interesting study in which they provide evidence that non-metricity $f(Q)$ gravity could challange $\Lambda$CDM model \cite{FKA}.

The article presented here is organized as: In Sec \ref{sec2}, we introduce the basic formulations governing the dynamics of $F(Q)$ gravity. In Sec \ref{sec3}, flat FLRW universe in $F(Q)$ gravity along with a dark energy component is presented. In Sec \ref{sec4}, we assume a cosmological $F(Q)$ model and derive the expressions for density, equation of state (EoS), and the deceleration parameter. Then we estimate the values of model parameters that would be in agreement with the observations. Further in Secs. \ref{sec5},\ref{sec6}, and \ref{sec7} we discuss the different behaviors of our cosmological model depending upon the power $n$ in the $F(Q)$ function. Finally, we present our outcomes in Sec \ref{sec8}.

\section{Fundamental Formulations in $F(Q)$ Gravity Theory}\label{sec2}

A metric-affine spacetime furnished with a general connection $X^{\alpha}_{\mu\nu}$ which establishes the notion of covariant derivatives and a metric tensor $g_{\mu\nu}$ that consists angles and distances. This general affine connection  $X^{\alpha}_{\mu\nu}$ can be decomposed into three parts \cite{Tom},

\begin{equation}\label{2a}
	X^\alpha_{\ \mu\nu}=\Gamma^\alpha_{\ \mu\nu}+K^\alpha_{\ \mu\nu}+L^\alpha_{\ \mu\nu},
\end{equation}

with the Levi-Civita connection of the metric tensor $g_{\mu\nu}$, 

\begin{equation}\label{2b}
	\Gamma^\alpha_{\ \mu\nu}\equiv\frac{1}{2}g^{\alpha\lambda}(g_{\mu\lambda,\nu}+g_{\lambda\nu,\mu}-g_{\mu\nu,\lambda})
\end{equation}

and the contortion tensor,

\begin{equation}\label{2c}
K^\alpha_{\ \mu\nu}\equiv\frac{1}{2}(T^{\alpha}_{\ \mu\nu}+T_{\mu \ \nu}^{\ \alpha}+T_{\nu \ \mu}^{\ \alpha})
\end{equation}
	
and the  distortion tensor,	
	
\begin{equation}\label{2d}
L^\alpha_{\ \mu\nu}\equiv\frac{1}{2}(Q^{\alpha}_{\ \mu\nu}-Q_{\mu \ \nu}^{\ \alpha}-Q_{\nu \ \mu}^{\ \alpha})
\end{equation}	

The terms in equations \eqref{2c} and \eqref{2d} are known as the torsion tensor and the non-metricity tensor respectively and it is given as
 
\begin{equation}\label{2e}
 T^\alpha_{\ \mu\nu}\equiv X^\alpha_{\ \mu\nu}-X^\alpha_{\ \nu\mu}
\end{equation}
 and 
\begin{equation}\label{2f}
Q_{\alpha\mu\nu}\equiv\nabla_\alpha g_{\mu\nu}
\end{equation}

As mentioned in the Introduction, the connection is assumed to be torsionless and curvatureless under the present setting, so that it corresponds to pure coordinate transformation from the trivial connection as described in \cite{J.B.}. The components of the connection can be can be determined by
 
\begin{equation}\label{2g}
X^\alpha \,_{\mu \beta} = \frac{\partial x^\alpha}{\partial \xi^\rho} \partial_ \mu \partial_\beta \xi^\rho.
\end{equation}

Here, $\xi^\alpha = \xi^\alpha (x^\mu)$ is an invertible relation. As discussed in Ref. \cite{ADK}, it is always possible to choose a coordinate system so that the connection $ X^\alpha_{\ \mu\nu} $ becomes trivial. The same gauge choice is used in several studies of STEGR \cite{ADK-2,IM} known as coincident gauge and in this situation the covariant derivative $\nabla_{\alpha}$ reduces to the partial derivative $\partial_{\alpha} $. However, for any other gauge choice where this general connection does not vanish, the metric evolution will be affected and result in a completely different theory \cite{ND,GDV}. Hence in the coincident gauge coordinate , one can have

\begin{equation}\label{2h}
Q_{\alpha\mu\nu} = \partial_\alpha g_{\mu\nu}
\end{equation}

while for an arbitrary coordinate choice,

\begin{equation}\label{2i}
Q_{\alpha\mu\nu}= \partial_\alpha g_{\mu\nu} - 2 X^\lambda_{\alpha (\mu} g_{\nu)\lambda}. 
\end{equation}


The following action governs the gravitational interactions in $F(Q)$ gravity,
\begin{equation}\label{2j}
S= \int{\frac{1}{2}F(Q)\sqrt{-g}d^4x} + \int{L_m\sqrt{-g}d^4x}
\end{equation}

Here $F(Q)$ is an arbitrary function of the non-metricity scalar $Q$,  $g$ is the determinant of the metric tensor $g_{\mu\nu}$, and $L_m$ is the matter Lagrangian.

Now, due to symmetricity of $g_{\mu\nu}$ there are only two independent traces acquired from the non-metricity term $Q_{\alpha\mu\nu}$ namely,
 
\begin{equation}\label{2k}
Q_\alpha = Q_\alpha\:^\mu\:_\mu \: and\:  \tilde{Q}_\alpha = Q^\mu\:_{\alpha\mu}
\end{equation}

In addition, the superpotential tensor is given by
\begin{equation}\label{2l}
4P^\lambda\:_{\mu\nu} = -Q^\lambda\:_{\mu\nu} + 2Q_{(\mu}\:^\lambda\:_{\nu)} + (Q^\lambda - \tilde{Q}^\lambda) g_{\mu\nu} - \delta^\lambda_{(\mu}Q_{\nu)}.
\end{equation}

Then the non-metricity scalar is given by \cite{LZ} 
\begin{equation}\label{2m}
Q = -Q_{\lambda\mu\nu}P^{\lambda\mu\nu}. 
\end{equation}

Moreover, the energy momentum tensor for the perfect fluid matter is given by

\begin{equation}\label{2n}
\mathcal{T}_{\mu\nu} = \frac{-2}{\sqrt{-g}} \frac{\delta(\sqrt{-g}L_m)}{\delta g^{\mu\nu}}
\end{equation}

Here, for simplicity we write $ F_Q = \frac{dF}{dQ} $ .

Now by varying the action \eqref{2j} with respect to the metric tensor,  we get the field equation representing the gravitational interactions in $F(Q)$ gravity as
\begin{widetext}
\begin{equation}\label{2p}
\frac{2}{\sqrt{-g}}\nabla_\lambda (\sqrt{-g}F_Q P^\lambda\:_{\mu\nu}) + \frac{1}{2}g_{\mu\nu}F+F_Q(P_{\mu\lambda\beta}Q_\nu\:^{\lambda\beta} - 2Q_{\lambda\beta\mu}P^{\lambda\beta}\:_\nu) = -T_{\mu\nu}.
\end{equation}
\end{widetext}

Furthermore, by varying the action for the connection, we have the following result,
\begin{equation}\label{2q}
\nabla_\mu \nabla_\nu (\sqrt{-g}F_Q P^{\mu\nu}\:_\lambda) =  0 
\end{equation}

\section{Flat FLRW Universe in $F(Q)$ Cosmology}\label{sec3}

Due to isotropy and homogeneity of the universe, we consider the following flat FLRW line element \cite{Ryden} in Cartesian coordinates, which is, as a matter of fact also a coincident gauge coordinates, therefore from now connection becomes trivial and metric is only a fundamental variable,

\begin{equation}\label{3a}
ds^2= -dt^2 + a^2(t)[dx^2+dy^2+dz^2]
\end{equation}

Here, $ a(t) $ is the scale factor . Then the non-metricity scalar corresponding to the line element \eqref{3a} is obtained as

\begin{equation}\label{3b}
 Q= 6H^2  
\end{equation}

The stress-energy momentum tensor describing the matter-content of the universe by isotropic pressure $p$ and energy density $\rho$ corresponding to the line element \eqref{3a} is 

\begin{equation}\label{3c}
\mathcal{T}_{\mu\nu}=(\rho+p)u_\mu u_\nu + pg_{\mu\nu}
\end{equation}

Here $u^\mu=(1,0,0,0)$ are the four velocity components characterizing the perfect cosmic fluid.\\

Then the Friedmann equations governing the dynamics of the universe for the function $F(Q)=-Q+f(Q)$, are 

\begin{equation}\label{3d}
f+Q-2Qf_Q = 2\rho
\end{equation}
and
\begin{equation}\label{3e}
\dot{H}=\frac{\rho+p}{2 \left(-1+f_Q+2Qf_{QQ} \right)} 
\end{equation}  

One can retrieve the Friedmann equations of General Relativity (i.e. STEGR limit) for the function $ F(Q)=-Q $.\\

Now, one can get the matter conservation equation
by taking trace of the field equation,
\begin{equation}\label{3f}
\dot{\rho} + 3H\left(\rho+p\right)=0
\end{equation}

The equation of state that relates the matter-energy density and the normal pressure of the cosmic fluid content is given as \cite{Som}
\begin{equation}\label{3g}
p=\omega \rho
\end{equation}

Here $\omega$ is constant called equation of state (EoS) parameter.

The equation \eqref{3f} can be solved by using equation \eqref{3g}, to give 

\begin{equation}\label{3h}
\rho=\rho_0 a(t)^{-3\left(1+\omega \right)}
\end{equation}

These equations \eqref{3d} and \eqref{3e} can be interpreted as symmetric teleparallel equivalent to GR (STG) cosmology with an additional component arising due to non-metricity of space-time which behaves like dark energy fluid part. These dark energy fluid components evolving due to non-metricity are defined by 

\begin{equation}\label{3i}
\rho_{DE}=-\frac{f}{2}+Qf_Q
\end{equation}

and 

\begin{equation}\label{3j}
p_{DE}= -\rho_{DE} - 2\dot{H} \left(f_Q+2Qf_{QQ} \right)
\end{equation}

Now, the dimensionless density parameter corresponding to the dark energy fluid part is defined as

\begin{equation}\label{3k}
\Omega_{DE}= \frac{\rho_{DE}}{3H^2}
\end{equation}

In addition, the EoS parameter that relates the dark energy density and its pressure is given by

\begin{equation}\label{3l}
\omega_{DE}= \frac{p_{DE}}{\rho_{DE}} = -1 + 4\dot{H} \left( \frac{f_Q+2Qf_{QQ}}{f-2Qf_Q} \right)
\end{equation}

Using \eqref{3d} and \eqref{3e}, we get

\begin{equation}\label{3m}
\omega_{DE} = -1 + \left(1+\omega\right) \frac{\left(f+Q-2Qf_Q\right) \left(f_Q+2Qf_{QQ} \right)}{\left(-1+f_Q+2Qf_{QQ} \right) \left(f-2Qf_Q \right)}
\end{equation}

Thus, the effective Friedmann equations along with a dark energy fluid part arising due to non-metricity reads

\begin{equation}\label{3n}
H^2=\frac{1}{3} \left[ \rho+\rho_{DE} \right]
\end{equation}

\begin{equation}\label{3o}
\dot{H}=-\frac{1}{2} \left[ \rho+p+\rho_{DE}+p_{DE} \right]
\end{equation}

Moreover, these dark energy fluid part satisfy the standard continuity equation

\begin{equation}\label{3p}
\dot{\rho}_{DE} + 3H \left( \rho_{DE} + p_{DE} \right) = 0
\end{equation}

\section{Cosmological $F(Q)$ Model}\label{sec4}

For our analysis, we consider the following $f(Q)$ model which is a combination of a linear and a non-linear term of non-metricity scalar, 
\begin{equation}\label{4a}
f(Q)= \alpha Q + \beta Q^n
\end{equation}
 
Here $\alpha$, $\beta$ and $n\neq1$ are free model parameters.

This particular functional form was motivated by a polynomial form applied, for instance, in Reference \cite{HRK}.

Then for this specific $f(Q)$ cosmological model, we obtained a first-order differential equation for a universe consisting of non-relativistic pressureless matter,

\begin{widetext}
\begin{equation}\label{4b}
\dot{H} \left[ \alpha-1+ \beta n (2n-1)6^{n-1} H^{2n-2}   \right] + \frac{3}{2} H^2  \left[ \alpha-1- \beta (2n-1)6^{n-1} H^{2n-2} \right]=0
\end{equation}
\end{widetext}

By using equations \eqref{3i} and \eqref{4a}, we obtain

\begin{equation}\label{4c}
\rho_{DE}= 3\alpha H^2 + \beta \frac{(2n-1)}{2} 6^n H^{2n}
\end{equation}

Again, by using equations \eqref{3k} and \eqref{3m}, we obtain

\begin{equation}\label{4d}
\Omega_{DE}= \frac{\rho_{DE}}{3H^2} = \alpha + \beta (2n-1) 6^{n-1} H^{2n-2}
\end{equation}
 
and

\begin{widetext}
\begin{equation}\label{4e}
\omega_{DE}= -1+ \frac{\left( \alpha-1+ \beta (2n-1) 6^{n-1} H^{2n-2} \right) \left( \alpha+ \beta n (2n-1) 6^{n-1} H^{2n-2}  \right)}{\left( \alpha-1+ \beta n (2n-1) 6^{n-1} H^{2n-2}\right)\left( \alpha+ \beta (2n-1) 6^{n-1} H^{2n-2}\right)} 
\end{equation}
\end{widetext}

Thus, the effective EoS parameter for our model is 

\begin{equation}\label{4f}
w_{eff}= \frac{p_{eff}}{\rho_{eff}}=\frac{p_{DE}}{\rho + \rho_{DE}}
\end{equation}

where, $p_{eff}$ and $\rho_{eff}$ correspond to the total pressure and energy density of the universe. Then we have

\begin{widetext}
\begin{equation}\label{4g}
w_{eff} = - \left( \alpha+ \beta (2n-1) 6^{n-1} H^{2n-2}\right) +  \frac{\left( \alpha-1+ \beta (2n-1) 6^{n-1} H^{2n-2} \right) \left( \alpha+ \beta n (2n-1) 6^{n-1} H^{2n-2}  \right)}{\left( \alpha-1+ \beta n (2n-1) 6^{n-1} H^{2n-2}\right)} 
\end{equation}
\end{widetext}

Another key component that characterize the expansion phase of the universe is deceleration parameter which is defined as \cite{Muj}

\begin{equation}\label{4h}
q= \frac{1}{2} \left( 1+3\Omega_{DE} \omega_{DE} \right)
\end{equation}

Using \eqref{4d} and \eqref{4e}, we get

\begin{widetext}
\begin{equation}\label{4i}
q=\frac{1}{2} + \frac{3}{2} \bigl\{ - \left( \alpha+ \beta (2n-1) 6^{n-1} H^{2n-2} \right) +  \frac{\left( \alpha-1+ \beta (2n-1) 6^{n-1} H^{2n-2} \right) \left( \alpha+ \beta n (2n-1) 6^{n-1} H^{2n-2}  \right)}{\left( \alpha-1+ \beta n (2n-1) 6^{n-1} H^{2n-2}\right)}  \bigr\}
\end{equation}
\end{widetext}

Now, we rewrite the Friedmann equations \eqref{3d} and \eqref{3e} with the help of dimensionless matter density parameter as

\begin{equation}\label{4j}
H^2=\frac{1}{12(-1+f_Q)} \left[ -Q \left( \Omega +1 \right) + f \right]
\end{equation}

\begin{equation}\label{4k}
 \dot{H}= \frac{1}{4(-1+f_Q)} \left[ \Omega Q-4f_{QQ} \dot{Q} H \right]
\end{equation}

where $\Omega=\frac{\rho}{3H^2} \ $ 

Our aim is to estimate the value of parameters of a given $f(Q)$ model that would be in agreement with the recently observed values of the cosmographic parameters. By using equations \eqref{4j} and \eqref{4k}, we obtain the values of model parameters $\alpha$ and $\beta$ in terms of present value cosmographic parameters and the power of non-metricity term $n$, as

\begin{equation}\label{4l}
\alpha = 1+\frac{\Omega_0}{(1-n)} \left[ n- \frac{3}{2 \left(1+q_0 \right)} \right]
\end{equation}
and
\begin{equation}\label{4m}
\beta=\frac{\Omega_0}{(2n-1)(1-n)6^{n-1} H_0^{2n-2}} \left[  \frac{3}{2 \left(1+q_0 \right)}-1 \right]
\end{equation}

where $n\neq \frac{1}{2} \: $ and $\: n\neq 1$

Now, we solve the differential equation \eqref{4b} by the numerical algorithm with initial conditions as $H_0=67.9 \: km/s/Mpc$ , $q_0=-0.55$, $\Omega_0=0.303$ with present time $t_0=13.8$ Gyrs \cite{Planck}. We have obtained the different behaviors of solution of our model depending upon the choice of power $n$ of the non-metricity term.

\section{Qunitessence like behavior of $F(Q)$ gravity model}\label{sec5}

\subsection{Cosmological Parameters}

\begin{figure}[H]
\includegraphics[scale=0.525]{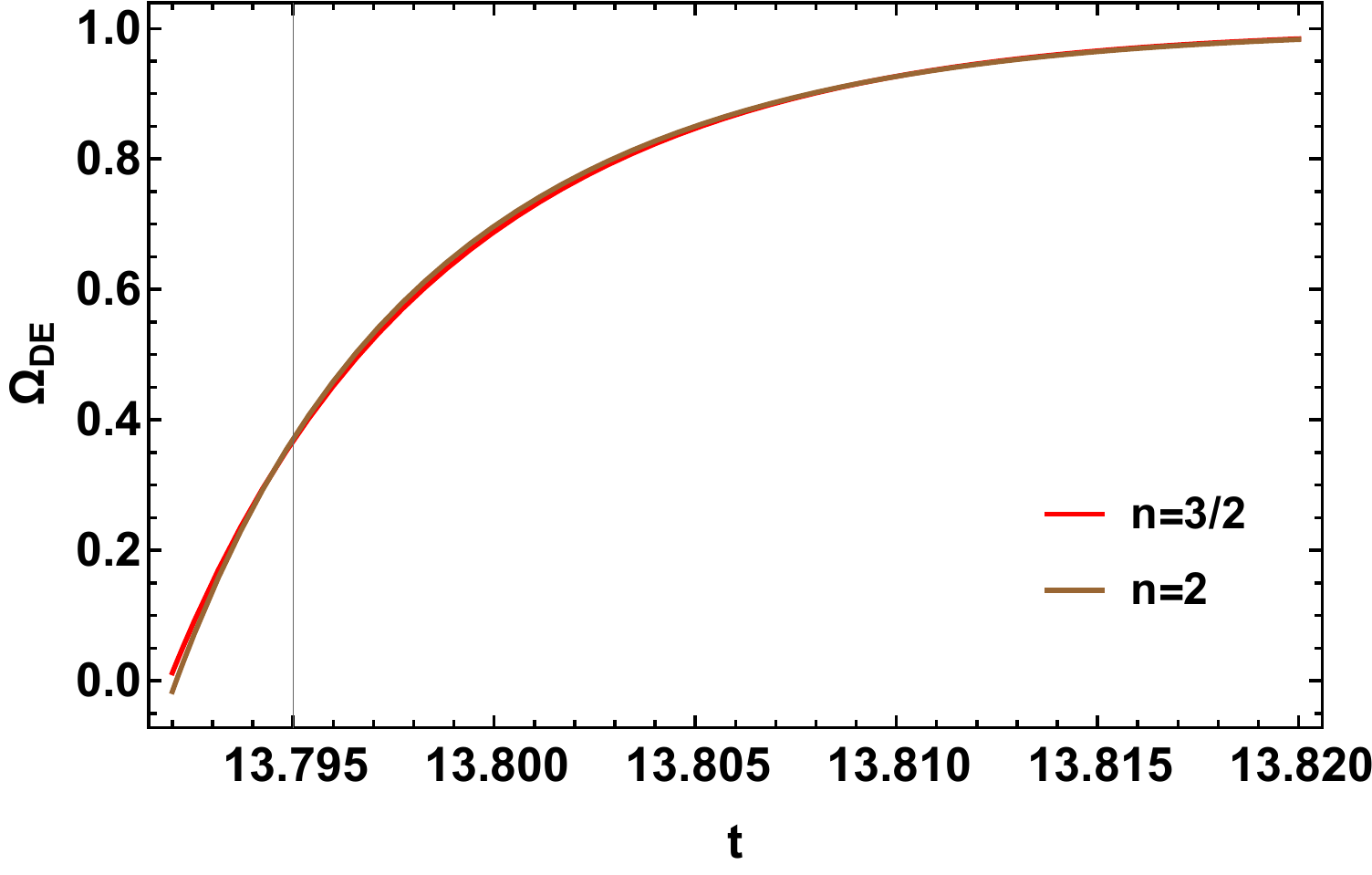}
\caption{Profile of the dimensionless density parameter for the dark energy component vs cosmic time t.}\label{f1}
\end{figure}

\begin{figure}[H]
\includegraphics[scale=0.51]{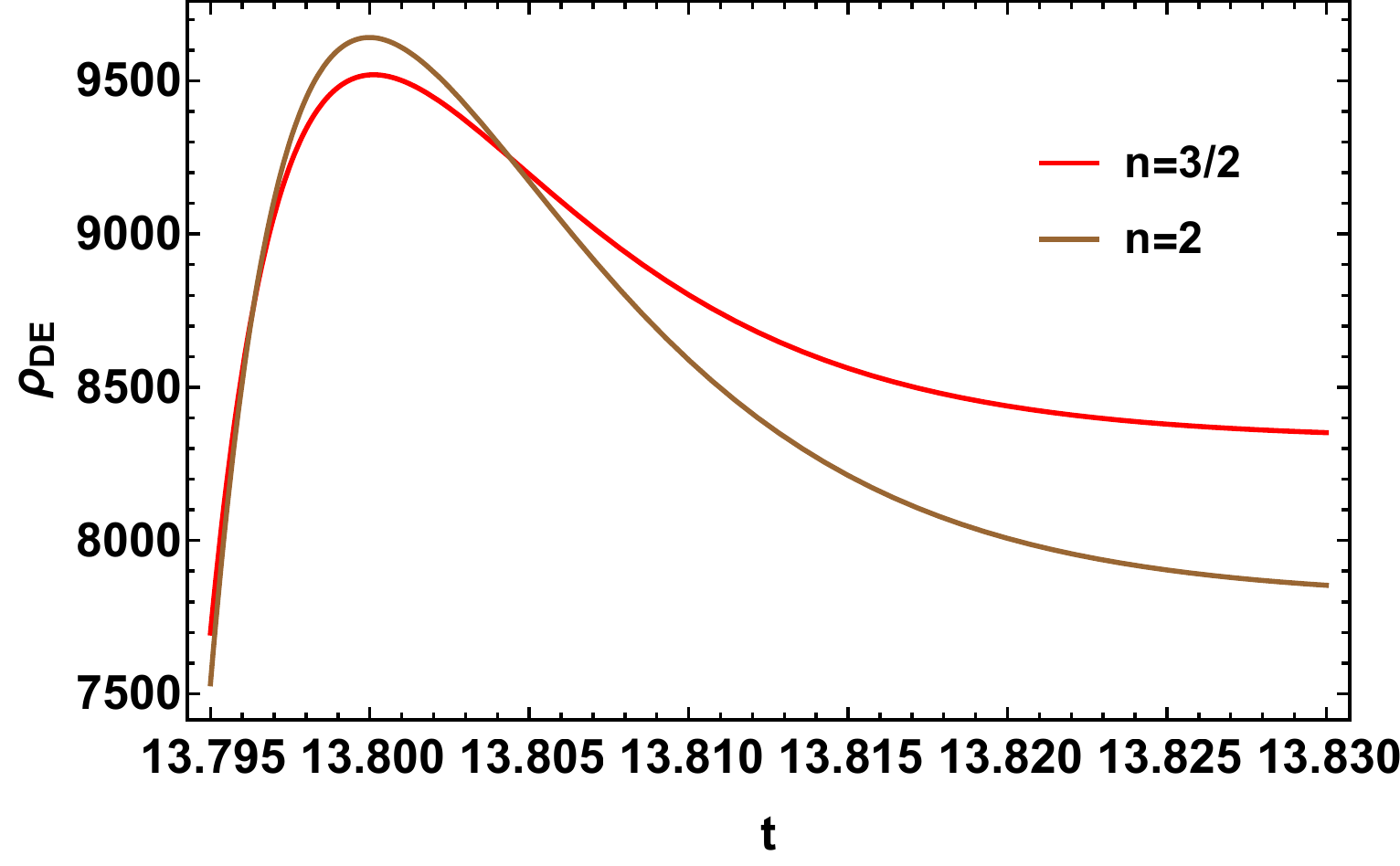}
\caption{Profile of the energy density of dark energy component vs cosmic time t. }\label{f2}
\end{figure}

\begin{figure}[H]
\includegraphics[scale=0.51]{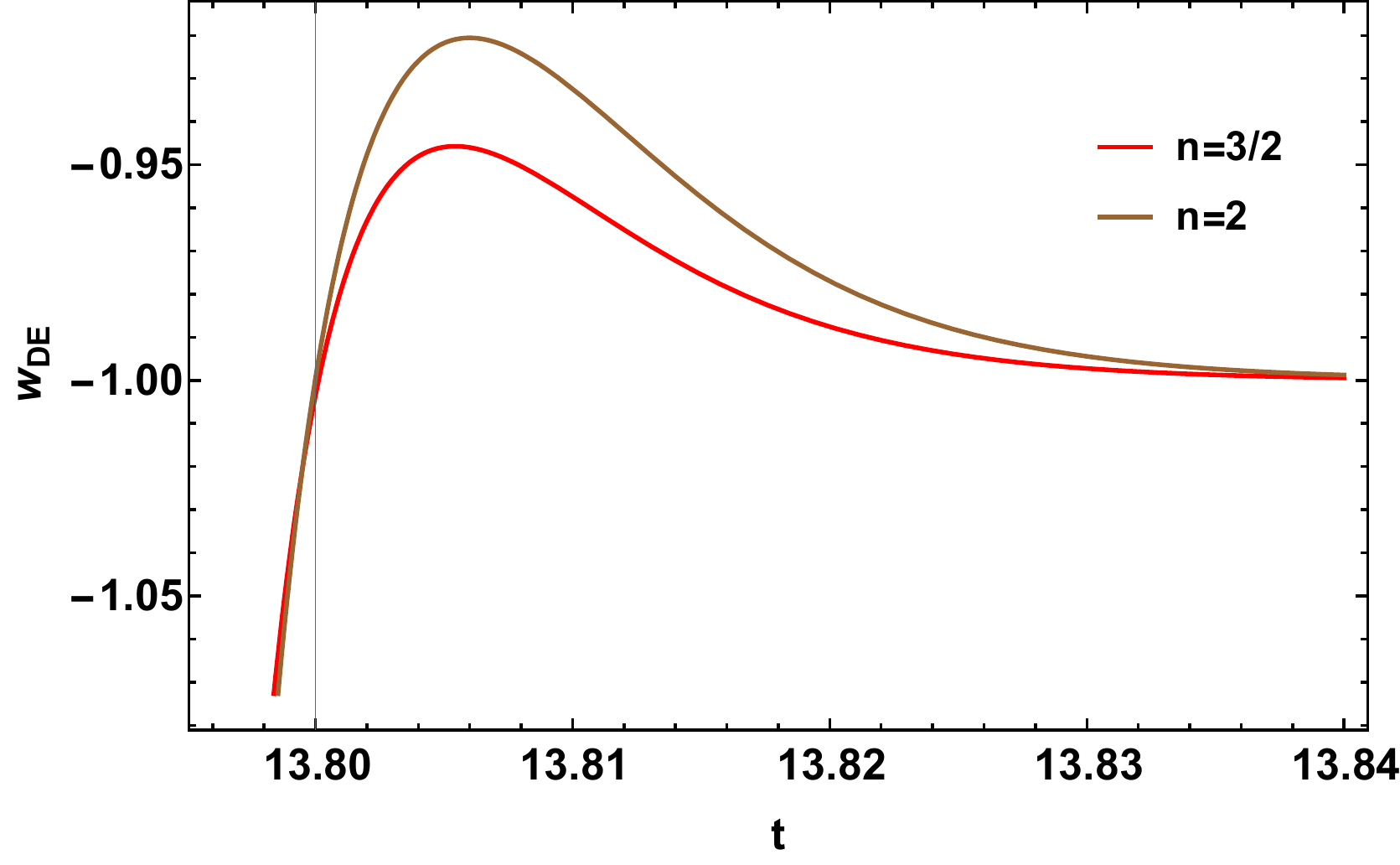}
\caption{Profile of the EoS parameter for the dark energy component vs cosmic time t. }\label{f3}
\end{figure}

\begin{figure}[H]
\includegraphics[scale=0.5]{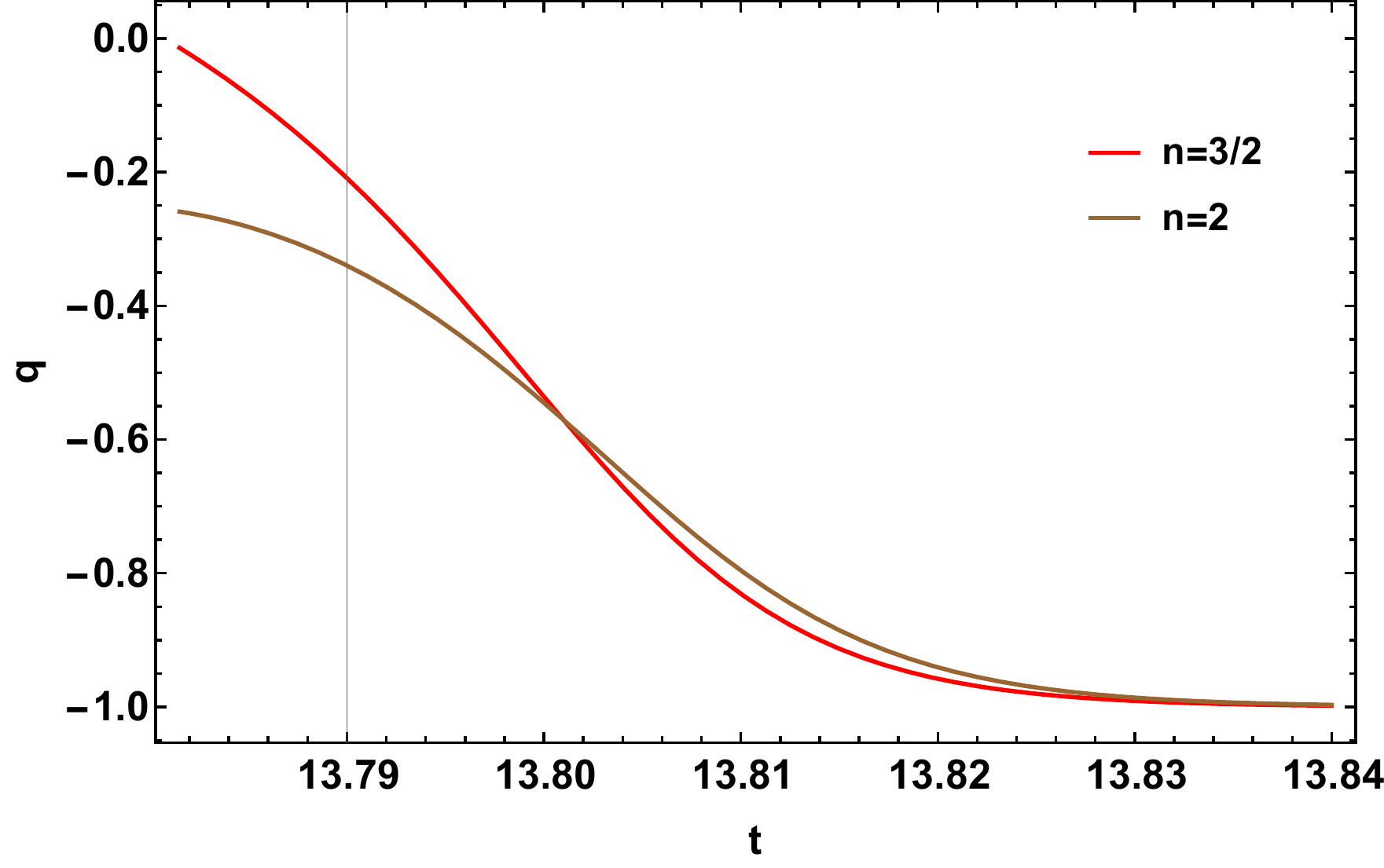}
\caption{Profile of the deceleration parameter vs cosmic time t.}\label{f4}
\end{figure}

\begin{figure}[H]
\includegraphics[scale=0.51]{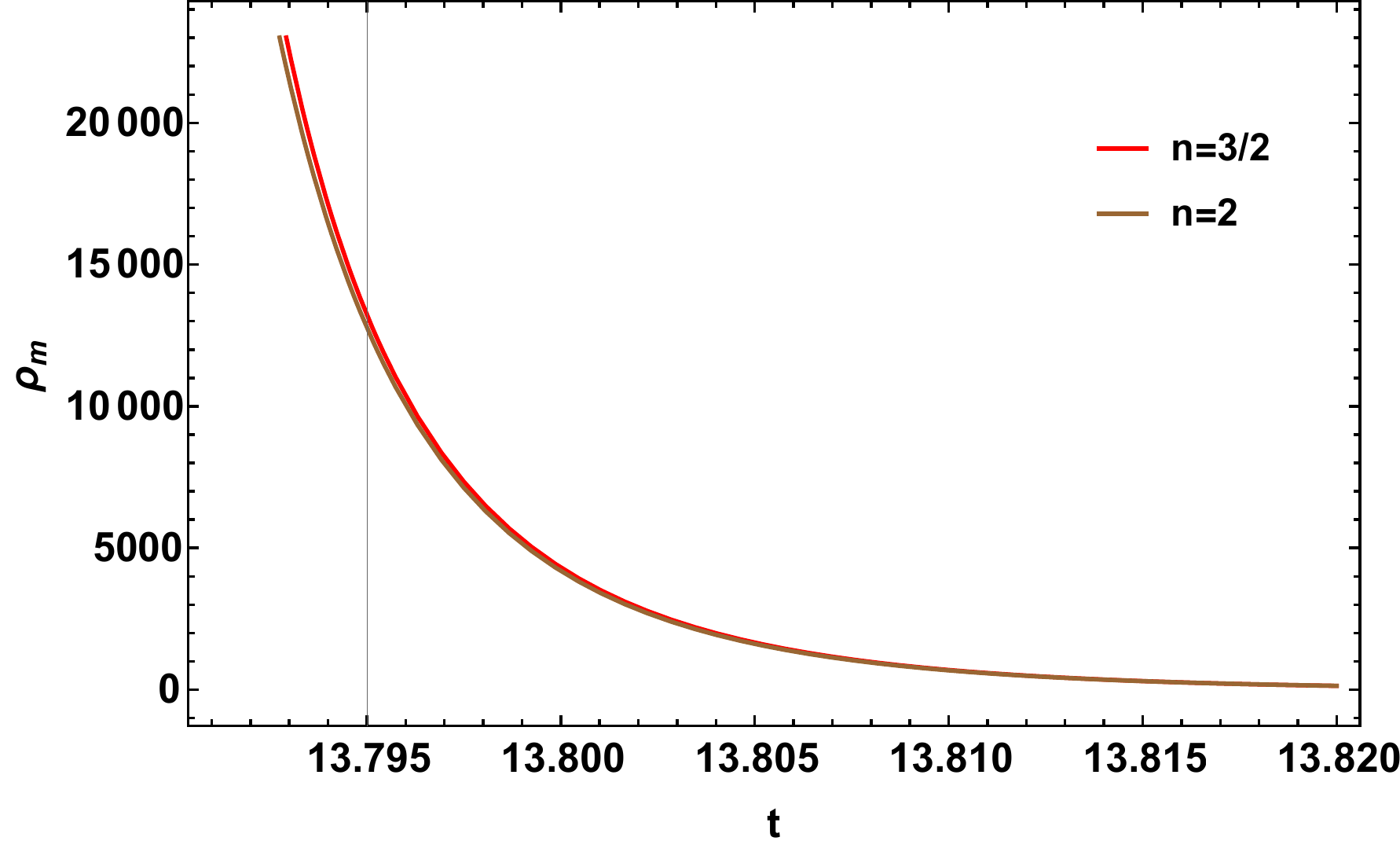}
\caption{Profile of the matter-energy density vs cosmic time t.}\label{f5}
\end{figure}

\begin{figure}[H]
\includegraphics[scale=0.5]{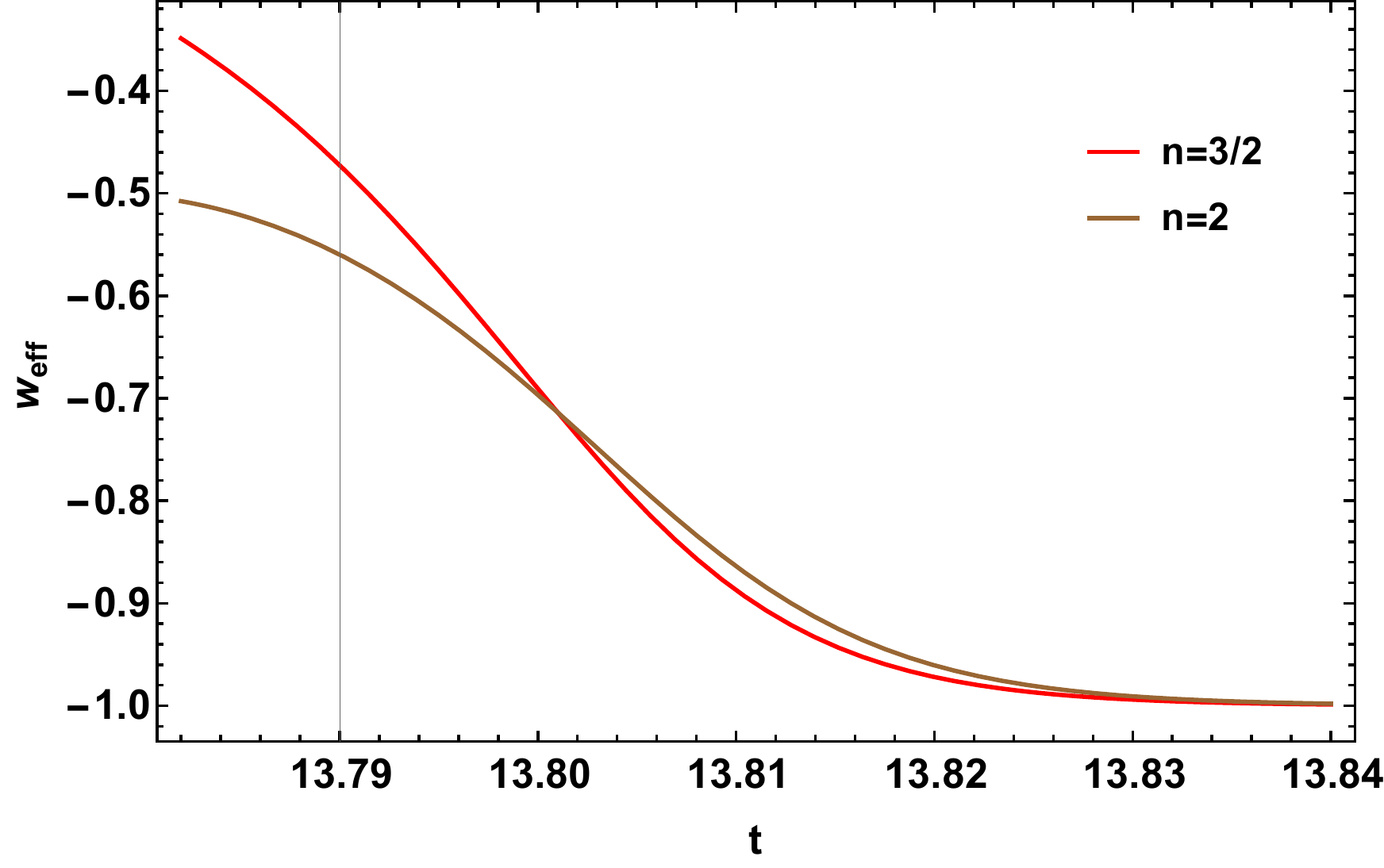}
\caption{Profile of the effective EoS parameter vs cosmic time t.}\label{f6}
\end{figure}


Form fig. \eqref{f1}, \eqref{f2} and \eqref{f5}, we found that the both matter-energy density and dark energy density of the universe decrease with cosmic time and matter energy density falls off to be zero while dark energy density becomes constant in the far future. Also, from fig. \eqref{f3} and \eqref{f6} it is clear that the  EoS parameter for both the dark energy and effective fluid converges to the $\Lambda$CDM equation of state and the dark energy evolving due to non-metricity shows quintessence like behavior \cite{KF,KC}. Further, fig \eqref{f4} indicates a transition from decelerating to accelerating phase of the universe in the recent past.

\subsection{Energy Conditions}

The energy conditions (ECs) have  a fundamental role to describe the geodesics of the universe and it can be derived from the Raychaudhuri equation \cite{EC}. The ECs are defined for the perfect fluid type effective matter content in $f(Q)$ gravity as\\

\begin{itemize}
\item Null energy condition (NEC) : $\rho_{eff}+p_{eff}\geq 0$  
\item Weak energy condition (WEC) : $\rho_{eff} \geq 0$ and  $\rho_{eff}+p_{eff}\geq 0$ 
\item Dominant energy condition (DEC) : $\rho_{eff} \pm p_{eff}\geq 0$ 
\item Strong energy condition (SEC) : $\rho_{eff}+ 3p_{eff}\geq 0$
\end{itemize}

\begin{figure}[H]
\includegraphics[scale=0.5]{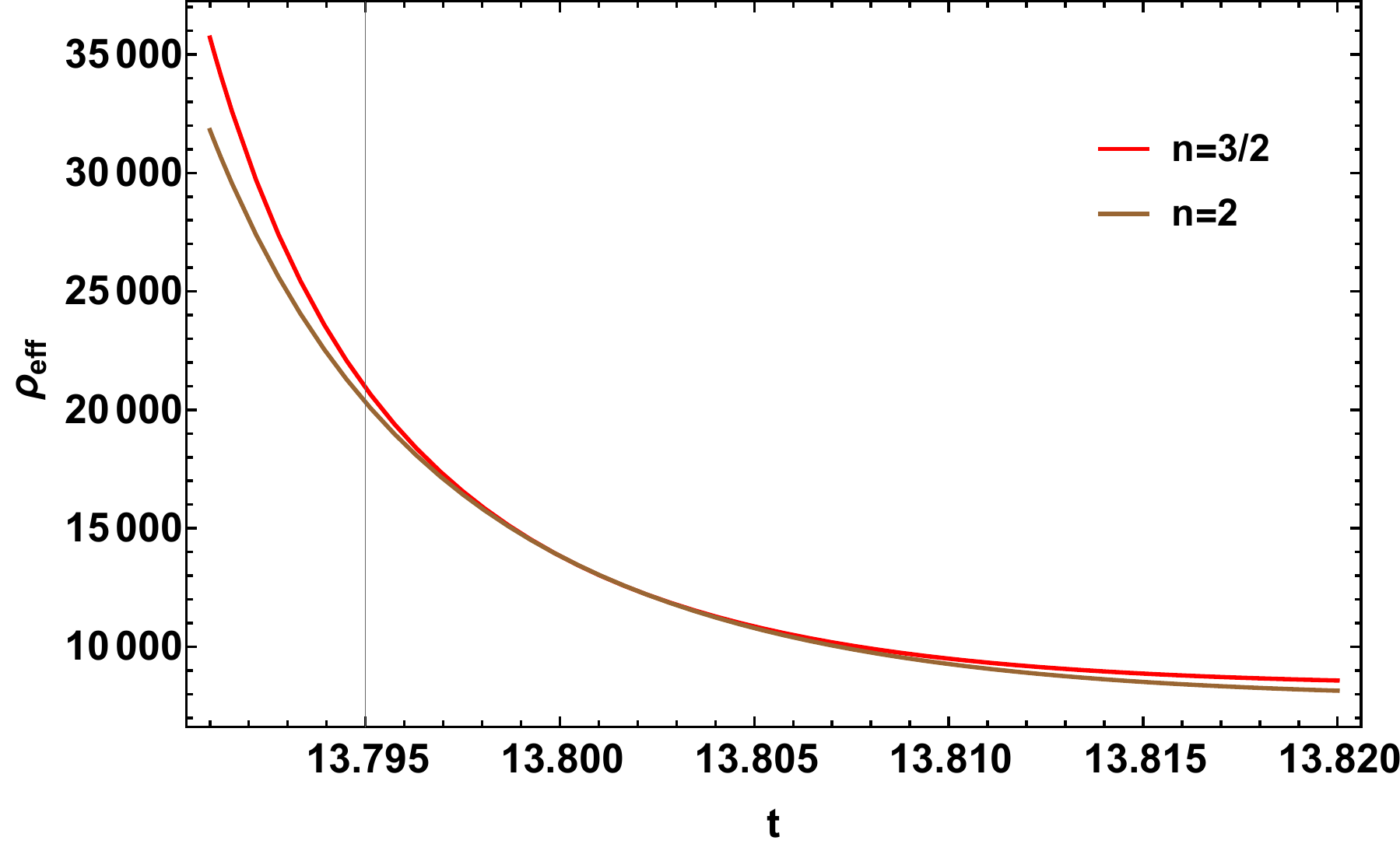}
\caption{Profile of the effective energy density vs cosmic time t.}\label{f7}
\end{figure}

\begin{figure}[H]
\includegraphics[scale=0.5]{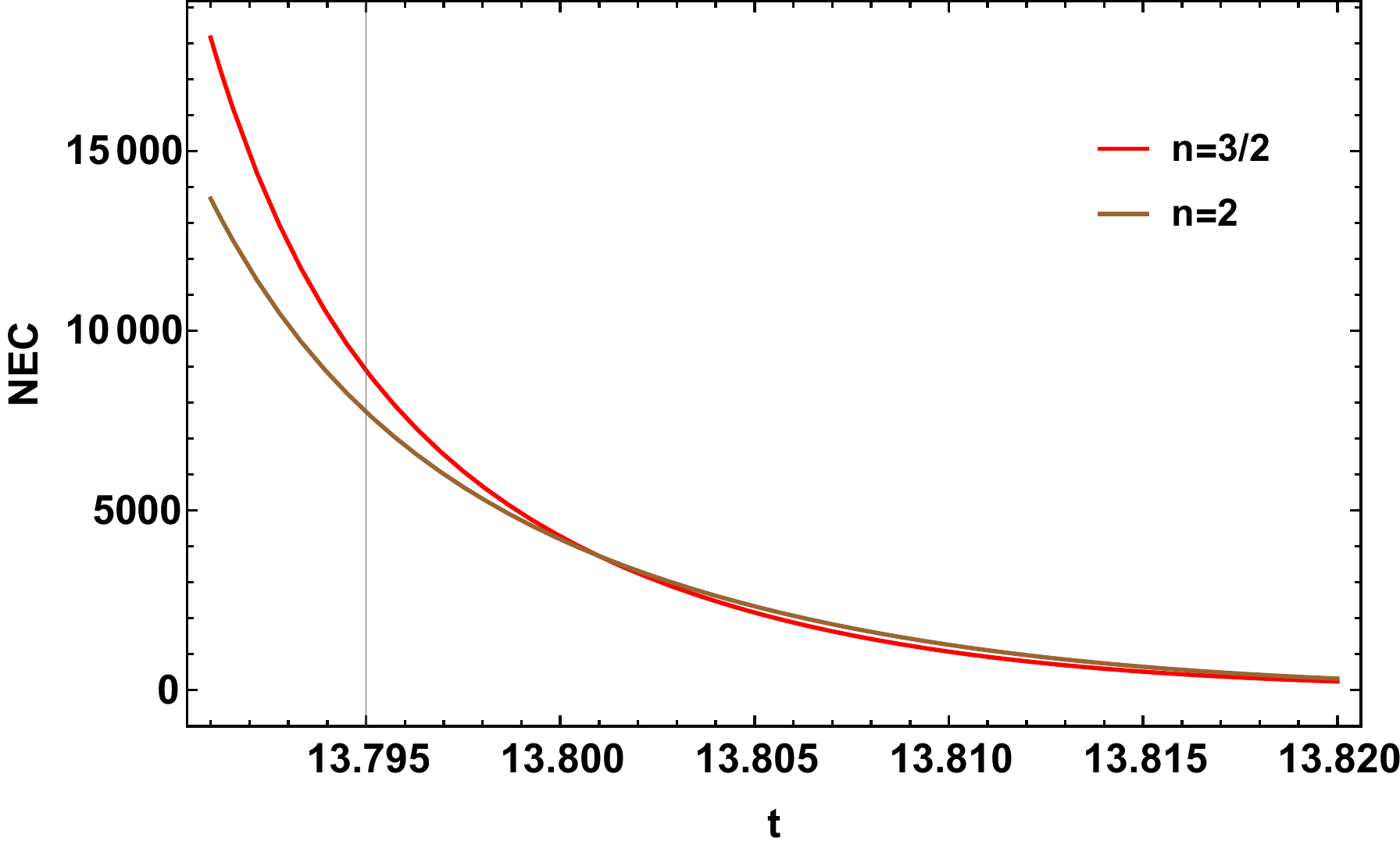}
\caption{Profile of the NEC vs cosmic time t.}\label{f8}
\end{figure}

\begin{figure}[H]
\includegraphics[scale=0.5]{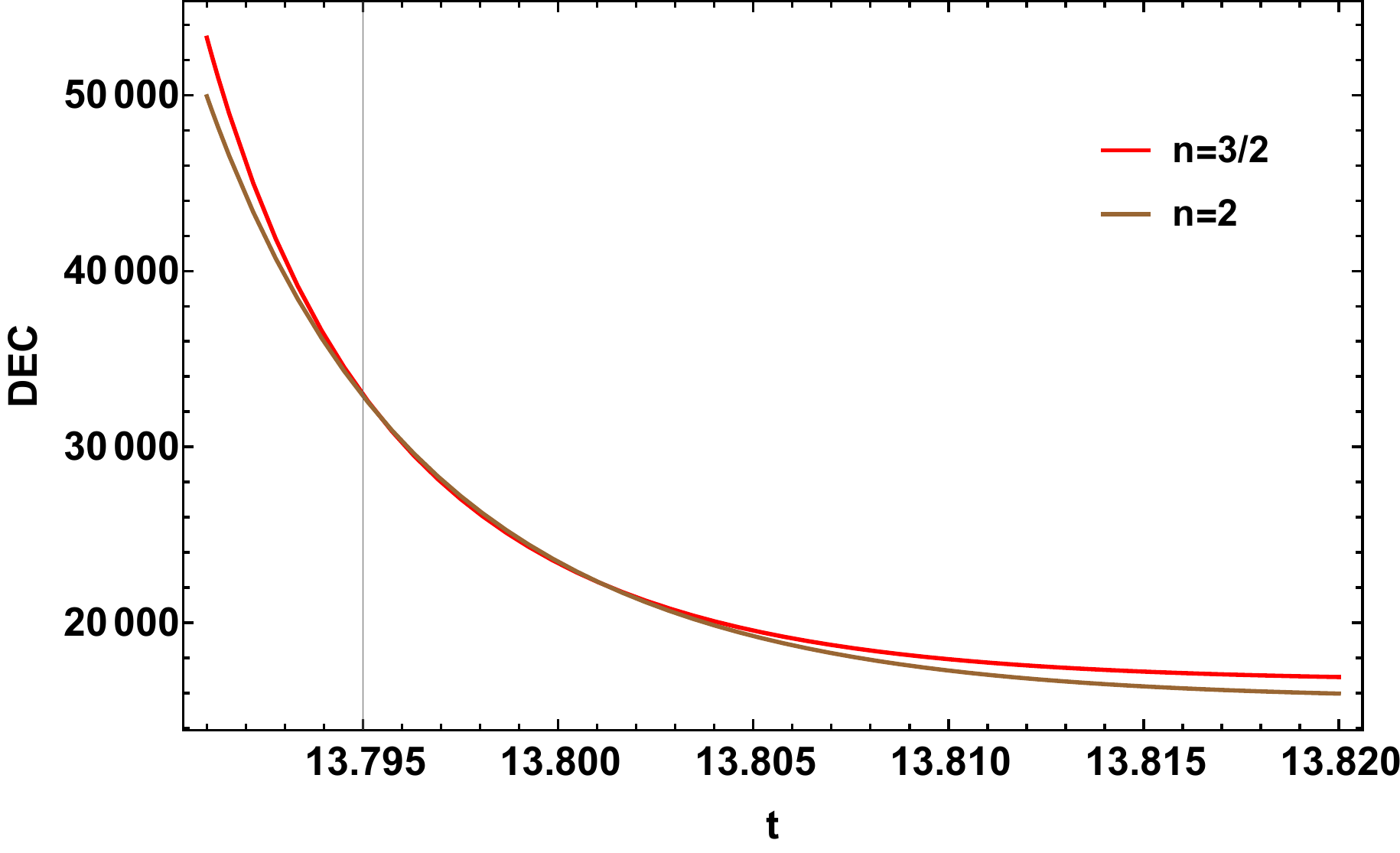}
\caption{Profile of the DEC vs cosmic time t.}\label{f9}
\end{figure}

\begin{figure}[H]
\includegraphics[scale=0.49]{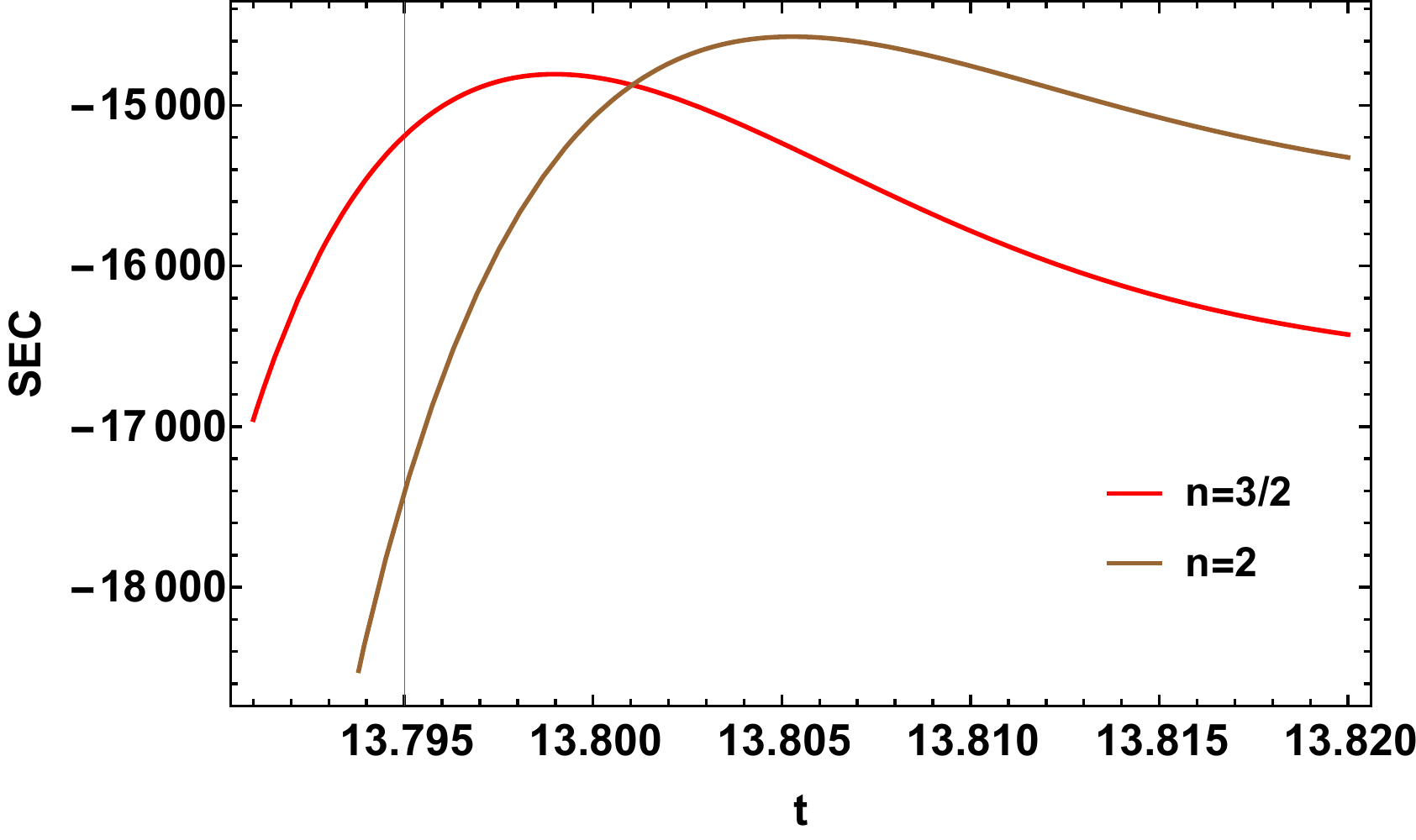}
\caption{Profile of the SEC vs cosmic time t.}\label{f10}
\end{figure}

From fig. \eqref{f7} it is clear that the effective energy density show positive behavior. Moreover from figs. \eqref{f8},\eqref{f9}, and \eqref{f10} we found that NEC, WEC, and DEC are satisfied while the SEC is violated. Violation of SEC depicts the acceleration phase of the universe. 

\section{Phantom like behavior of $F(Q)$ gravity model}\label{sec6}

\subsection{Cosmological Parameters}

\begin{figure}[H]
\includegraphics[scale=0.51]{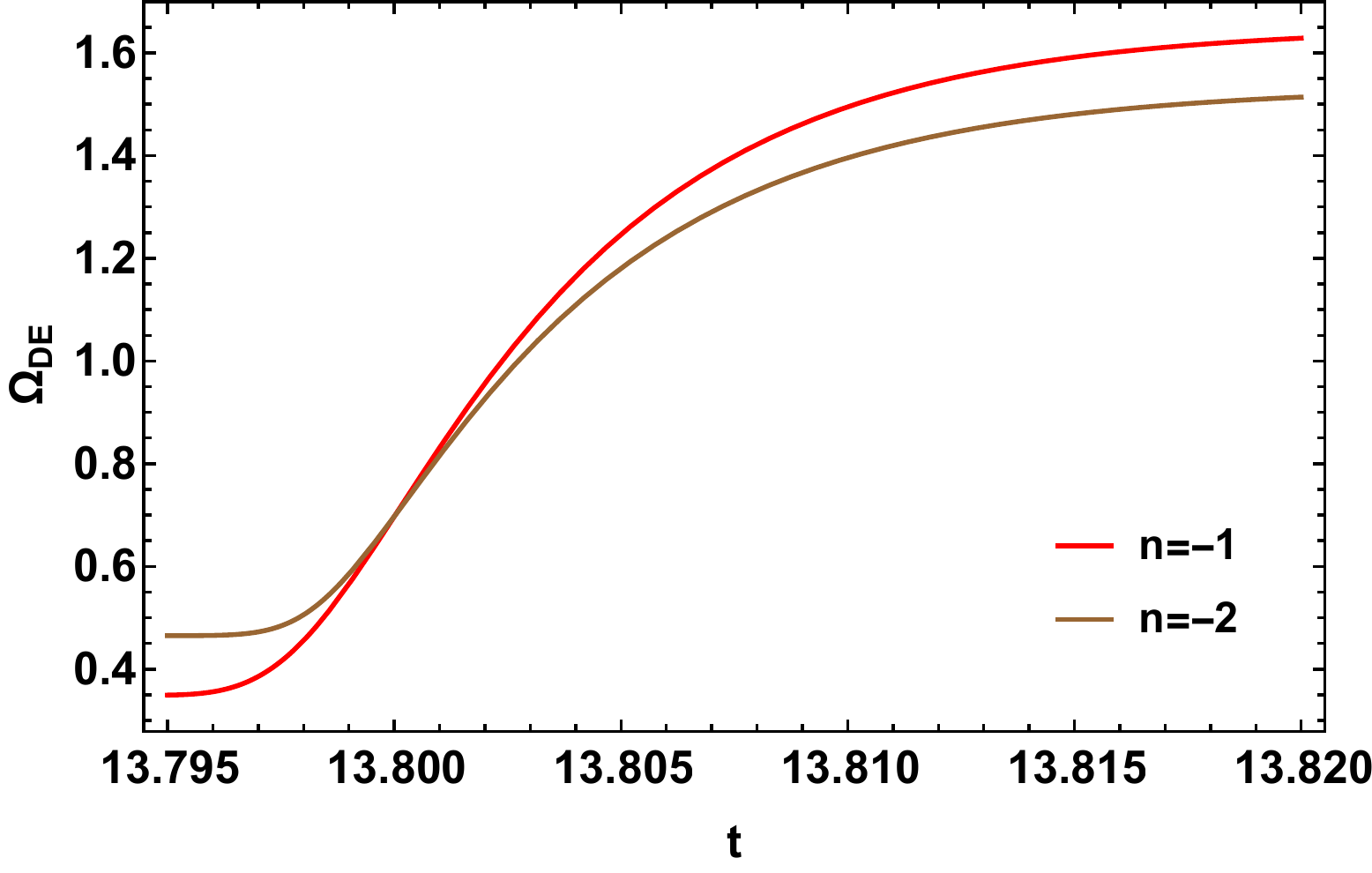}
\caption{Profile of the dimensionless density parameter for the dark energy component vs cosmic time t.}\label{f11}
\end{figure}

\begin{figure}[H]
\includegraphics[scale=0.51]{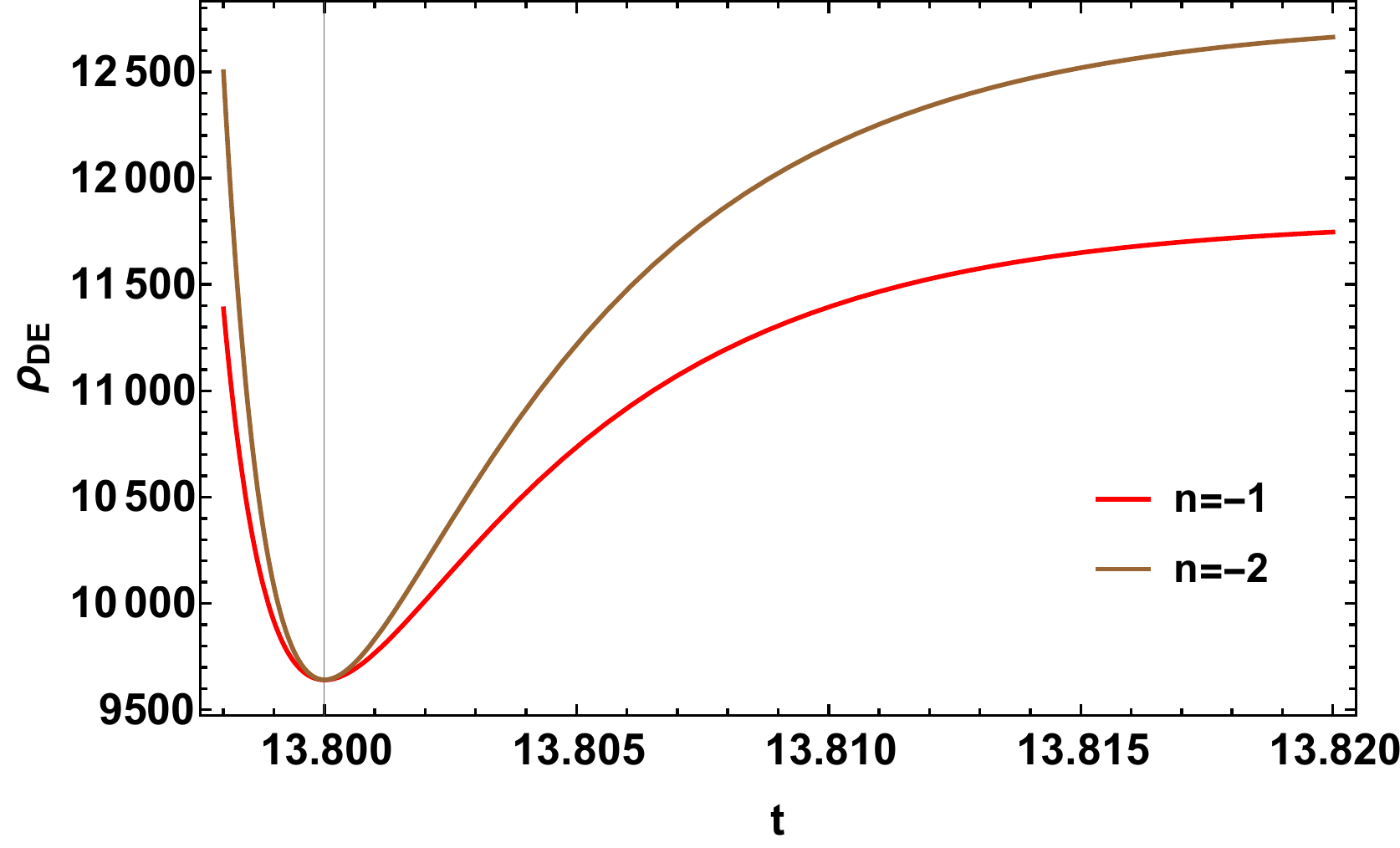}
\caption{Profile of the energy density of dark energy component vs cosmic time t. }\label{f12}
\end{figure}

\begin{figure}[H]
\includegraphics[scale=0.51]{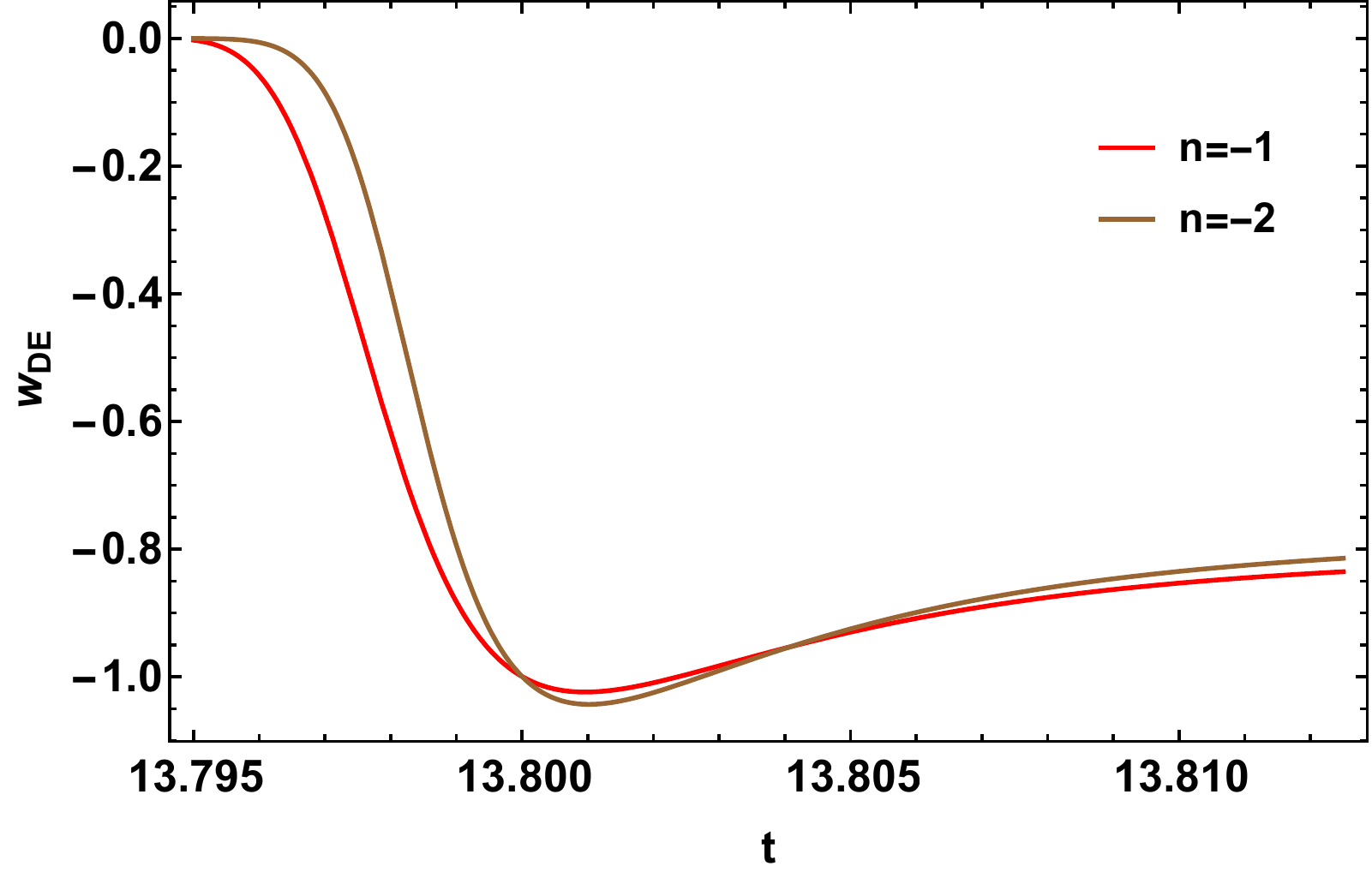}
\caption{Profile of the EoS parameter for the dark energy component vs cosmic time t. }\label{f13}
\end{figure}

\begin{figure}[H]
\includegraphics[scale=0.51]{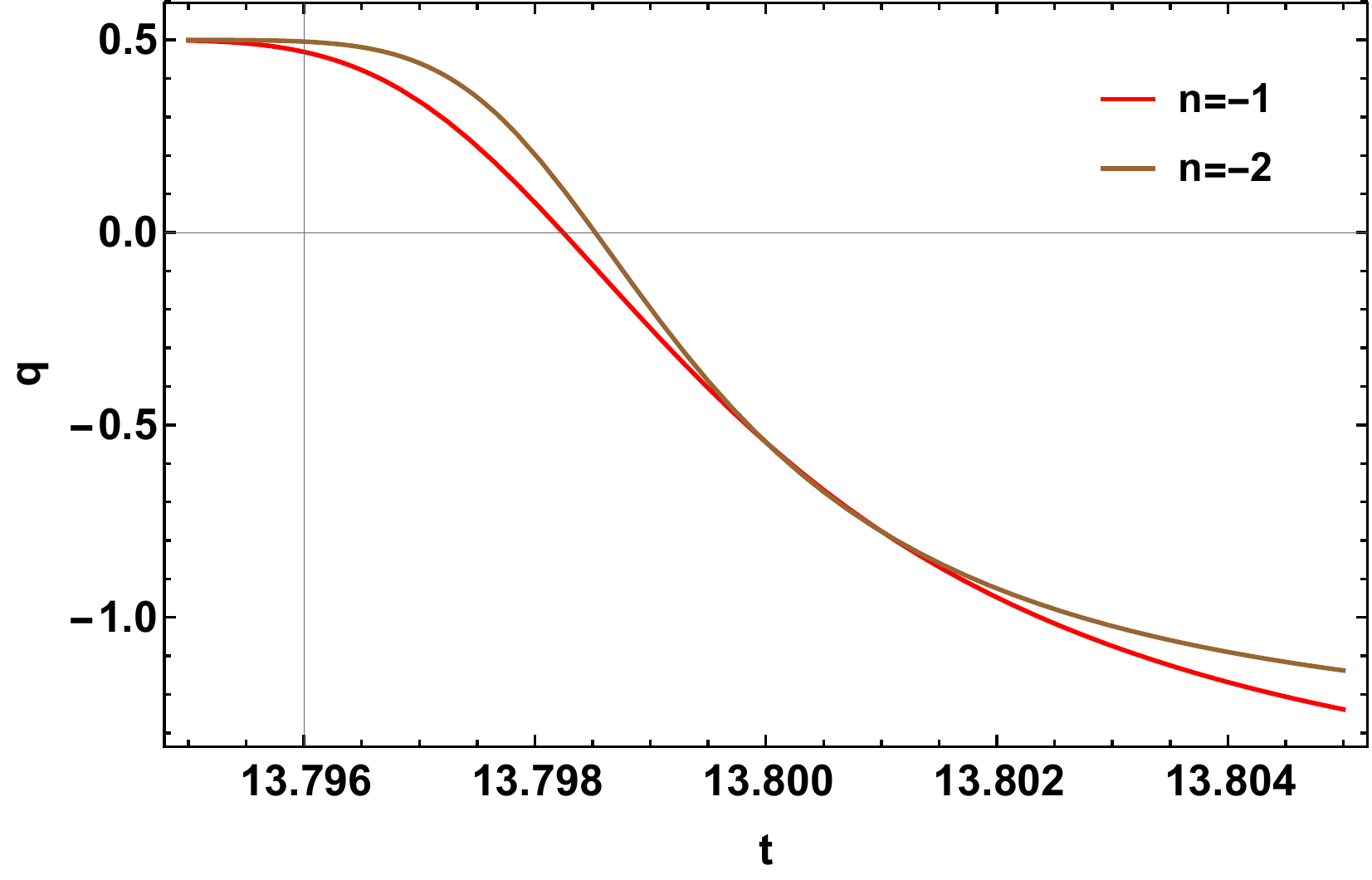}
\caption{Profile of the deceleration parameter vs cosmic time t.}\label{f14}
\end{figure}

\begin{figure}[H]
\includegraphics[scale=0.51]{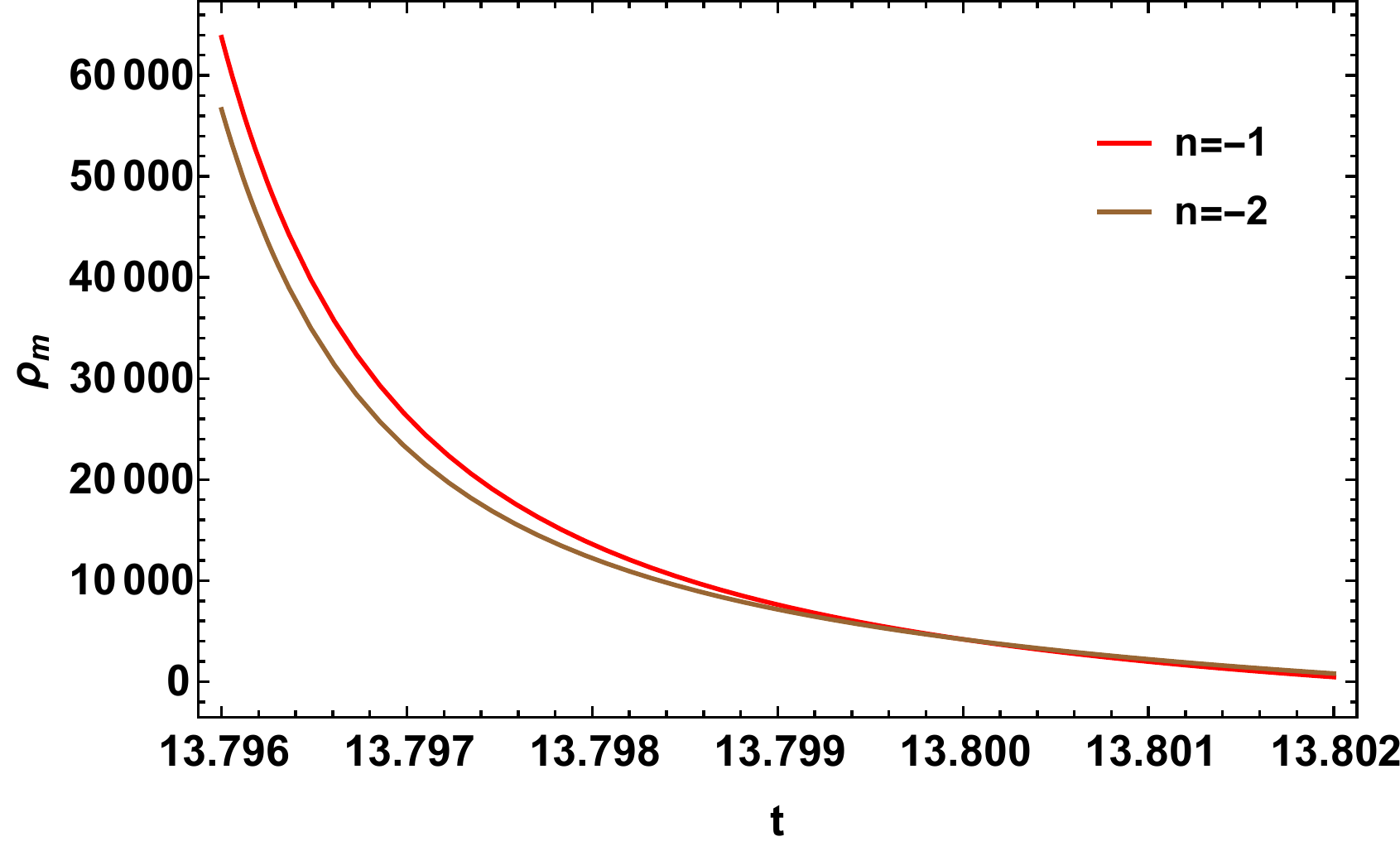}
\caption{Profile of the matter-energy density vs cosmic time t.}\label{f15}
\end{figure}

\begin{figure}[H]
\includegraphics[scale=0.51]{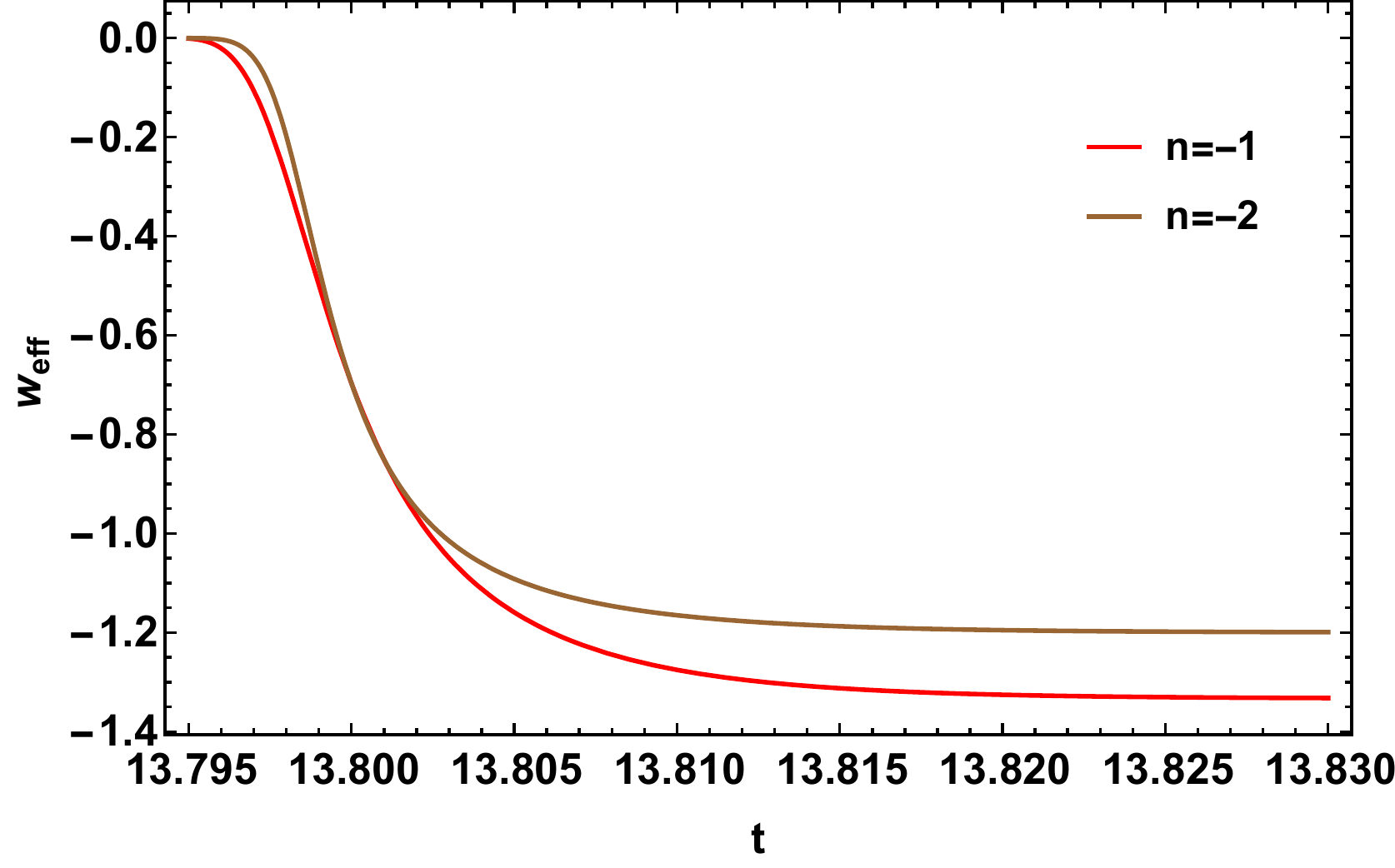}
\caption{Profile of the effective EoS parameter vs cosmic time t.}\label{f16}
\end{figure}

Form fig. \eqref{f11}, \eqref{f12} and \eqref{f15}, we found that the matter-energy density of the universe decrease with cosmic time while dark energy density increases with time. Moreover, the matter energy density falls off to be zero while dark energy density becomes constant in the far future.  Also, from fig. \eqref{f13} and \eqref{f16} we found that the EoS parameter for the dark energy fluid shows a transiting behavior from phantom to non-phantom phase while the effective EoS parameters crosses the $\Lambda$CDM line and then converges in phantom region. Hence, the dark energy evolving due to non-metricity shows phantom like behavior. Further, fig \eqref{f14} indicates a transition from decelerating to accelerating phase of the universe in the recent past.


\subsection{Energy Conditions}

\begin{figure}[H]
\includegraphics[scale=0.5]{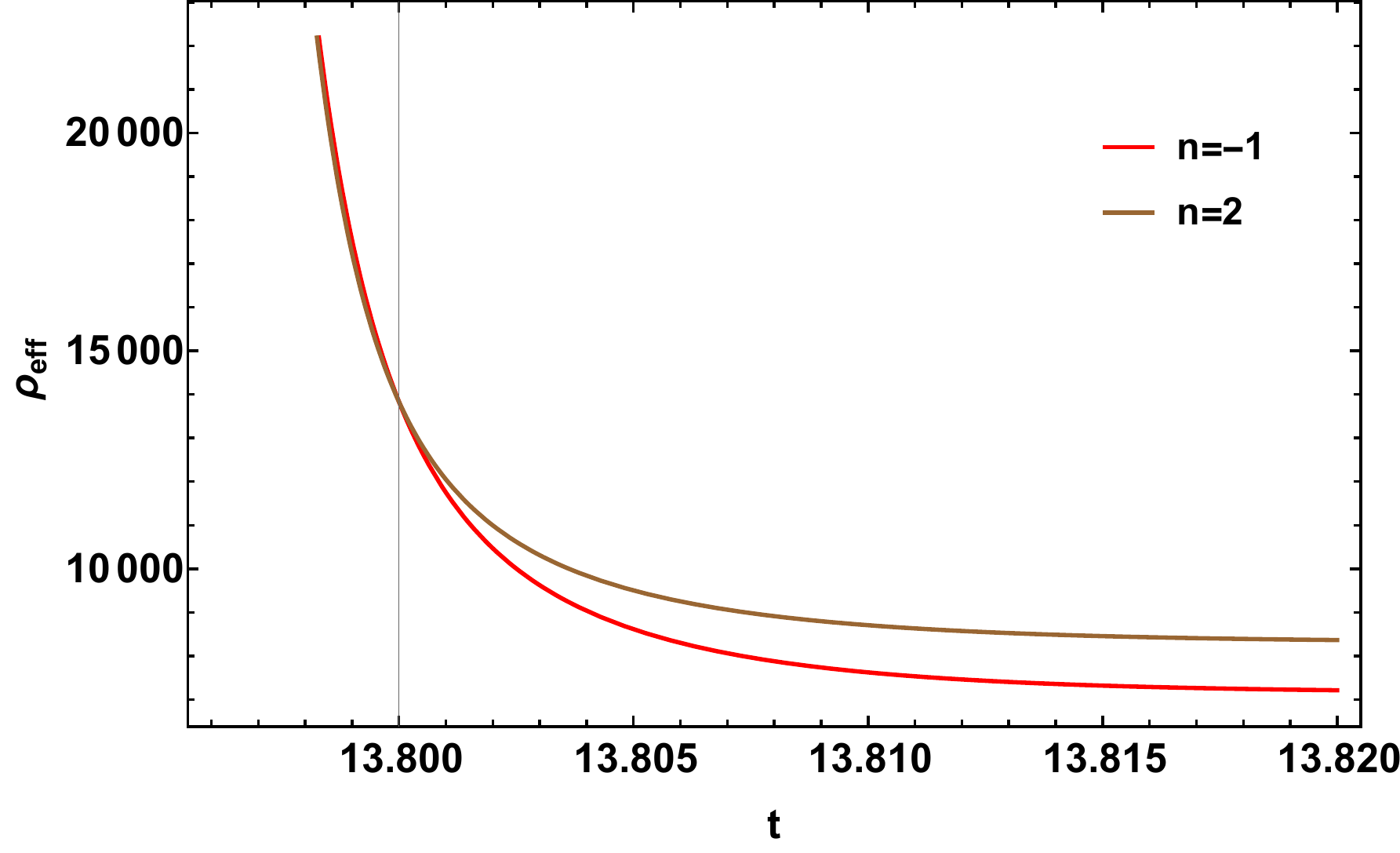}
\caption{Profile of the effective energy density vs cosmic time t.}\label{f17}
\end{figure}

\begin{figure}[H]
\includegraphics[scale=0.5]{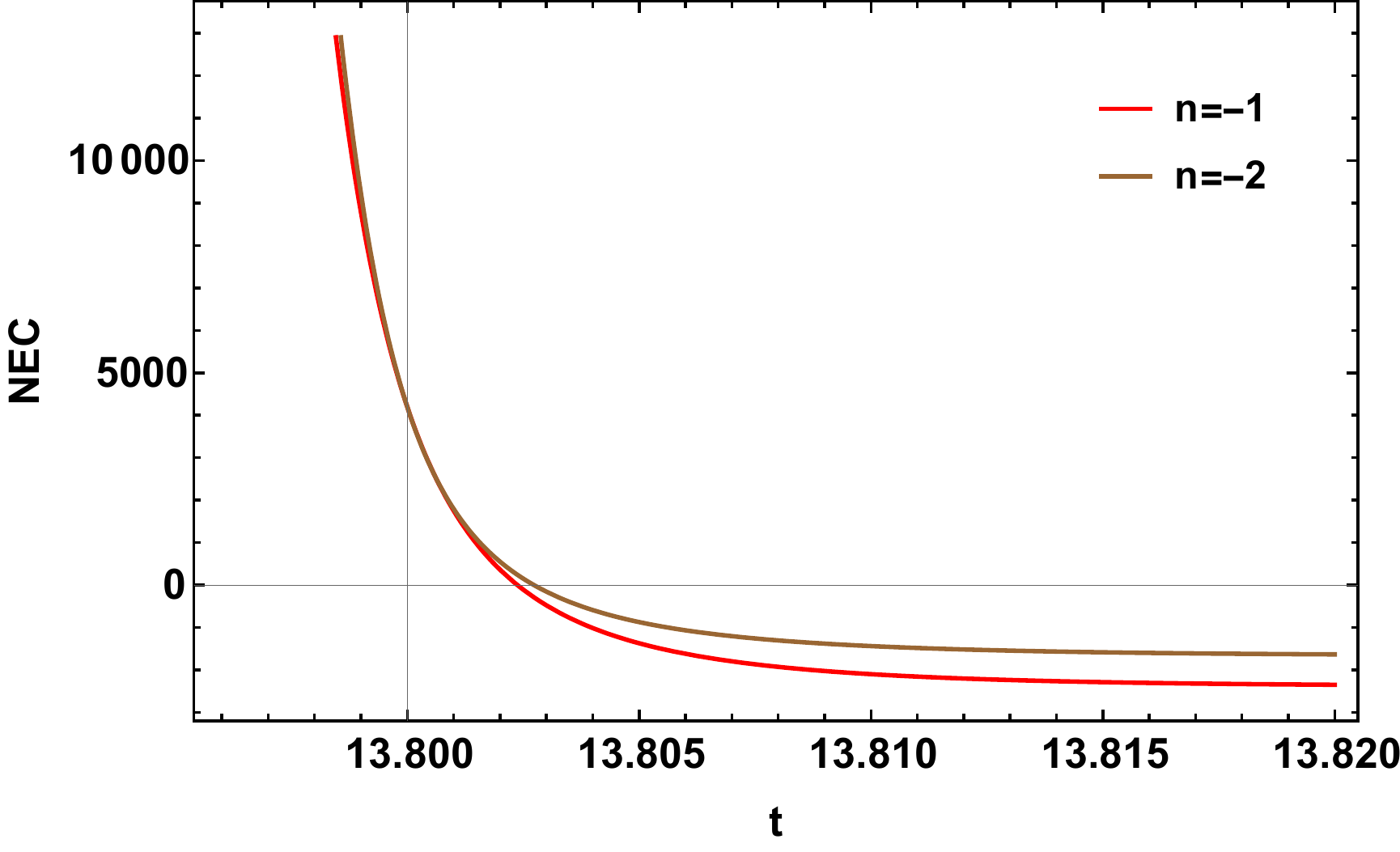}
\caption{Profile of the NEC vs cosmic time t.}\label{f18}
\end{figure}

\begin{figure}[H]
\includegraphics[scale=0.51]{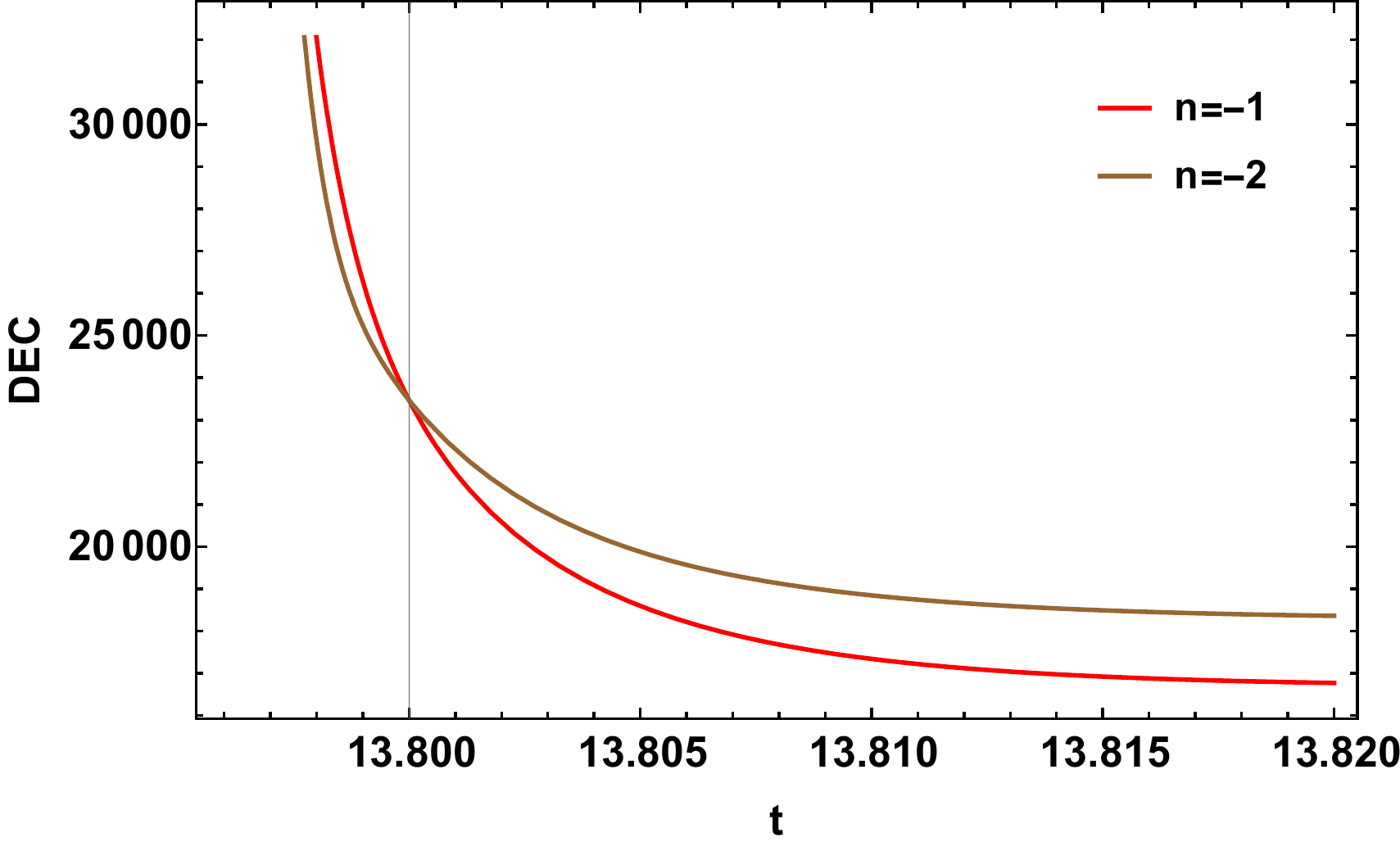}
\caption{Profile of the DEC vs cosmic time t.}\label{f19}
\end{figure}

\begin{figure}[H]
\includegraphics[scale=0.53]{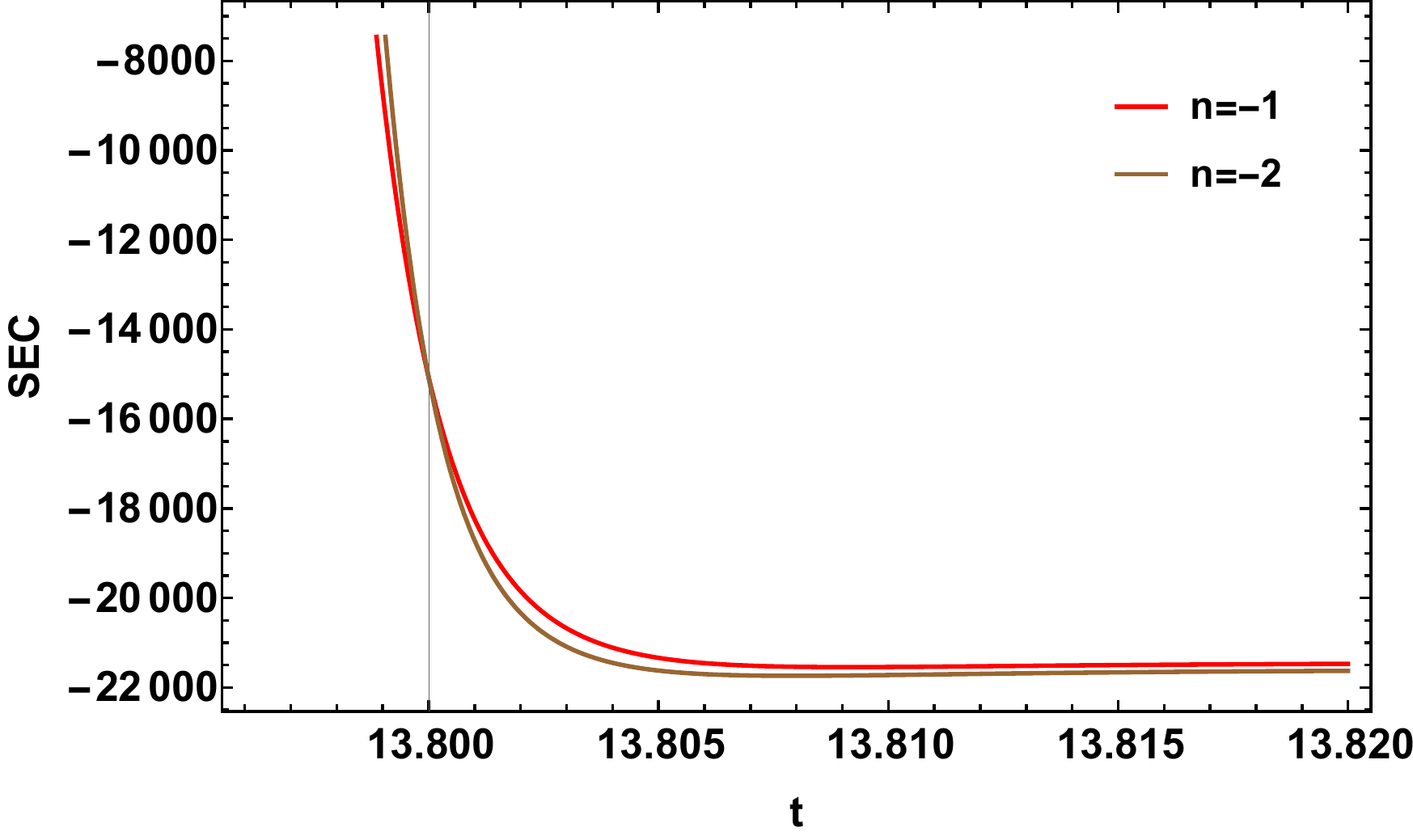}
\caption{Profile of the SEC vs cosmic time t.}\label{f20}
\end{figure}

From fig. \eqref{f17} it is clear that the effective energy density show positive behavior. Moreover from figs. \eqref{f18} and \eqref{f20} we found that NEC and SEC are violated. By the definition of energy conditions violation of NEC implies the violation of WEC and DEC. Thus all the ECs are violated in this case. Violation of SEC depicts the acceleration phase of the universe while the violation of NEC depicts the existence of exotic matter \cite{ZZ}. Thus the violation of ECs also shows that the dark energy evolving due to non-metricity behaves like phantom energy.

\section{$\Lambda$CDM like behavior of $F(Q)$ gravity model}\label{sec7}

Consider the case $\alpha=0 \:$, $\: n=0 \:$ and $\: \beta < 0 \:$ i.e,   $F(Q)=-Q+\beta$.\\
\\
In this case we have,\\
$\rho_{DE}= -\frac{\beta}{2} = constant \:$ , $\: p_{DE}= \frac{\beta}{2} = constant \:$ and $\: \omega_{DE}=-1 $.\\
Also, the density parameter and effective EoS parameter becomes 

\begin{equation}
\Omega_{DE}= - \frac{\beta}{6H^2}
\end{equation}
and
\begin{equation}
w_{eff} = \frac{\beta}{6H^2}
\end{equation}
and
\begin{equation}
q=\frac{1}{2} + \frac{\beta}{4H^2}
\end{equation}

\begin{figure}[H]
\includegraphics[scale=0.51]{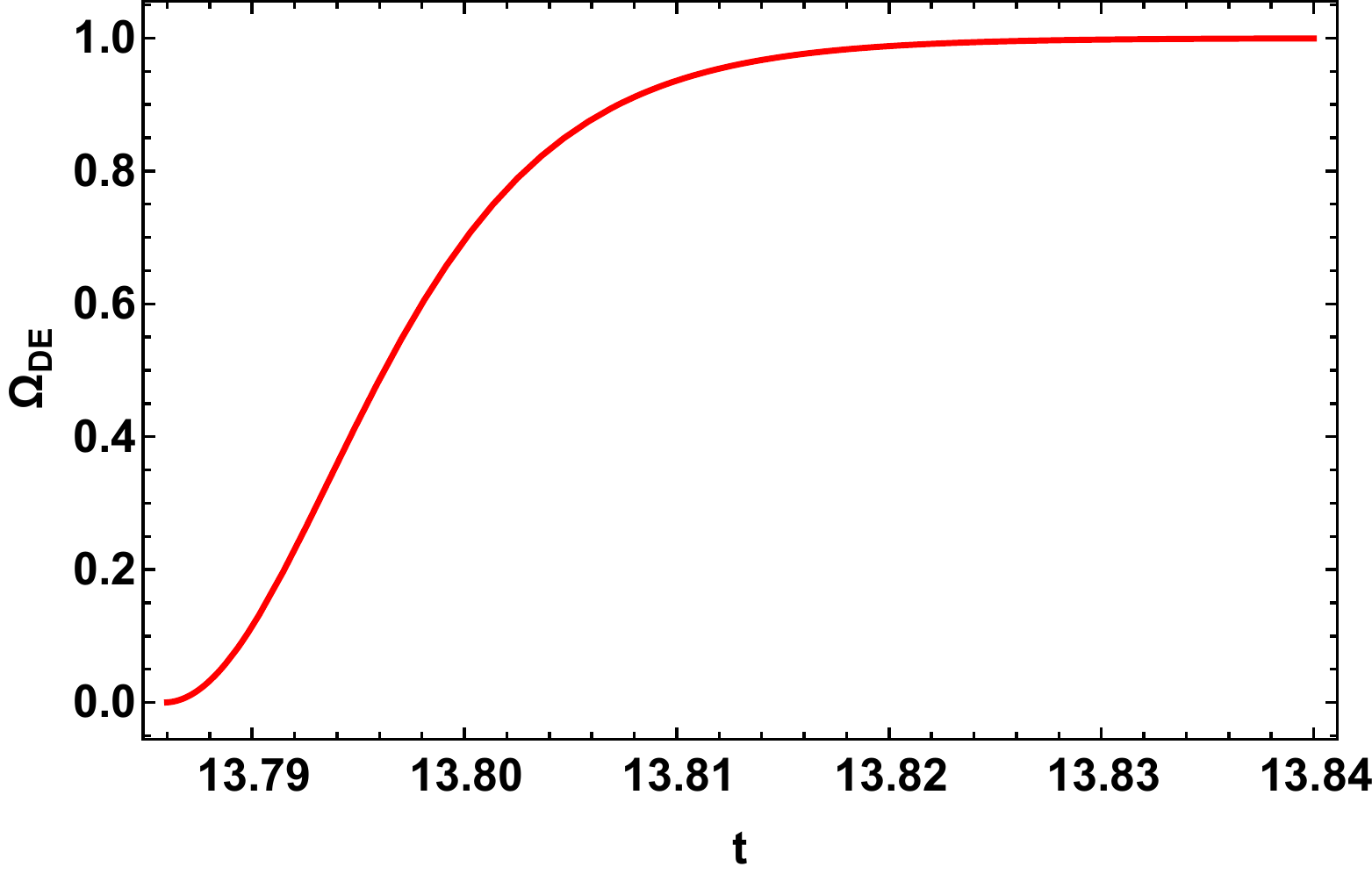}
\caption{Profile of the dimensionless density parameter for the dark energy component vs cosmic time t.}\label{f21}
\end{figure}

\begin{figure}[H]
\includegraphics[scale=0.51]{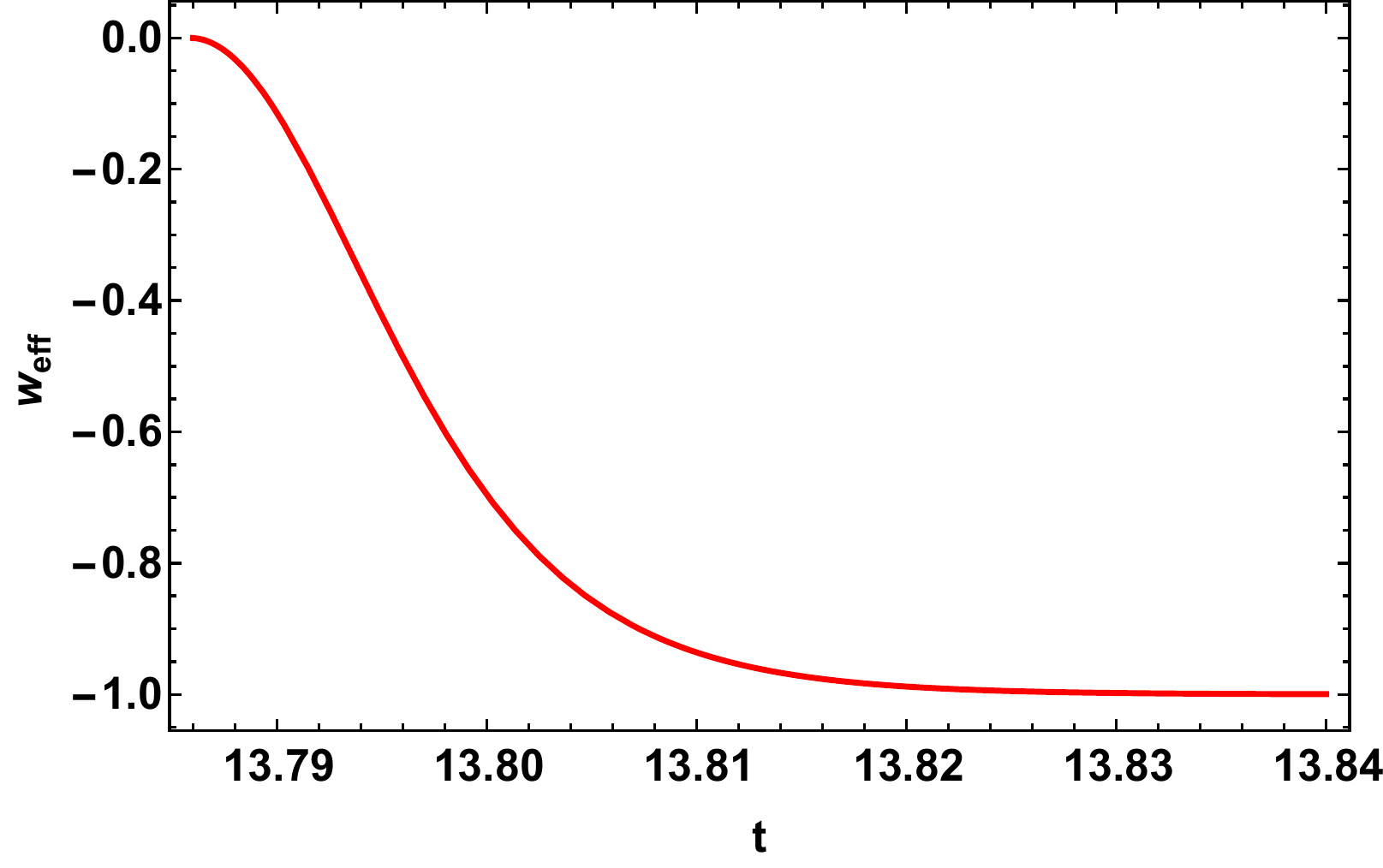}
\caption{Profile of the effective EoS parameter vs cosmic time t.}\label{f22}
\end{figure}

\begin{figure}[H]
\includegraphics[scale=0.51]{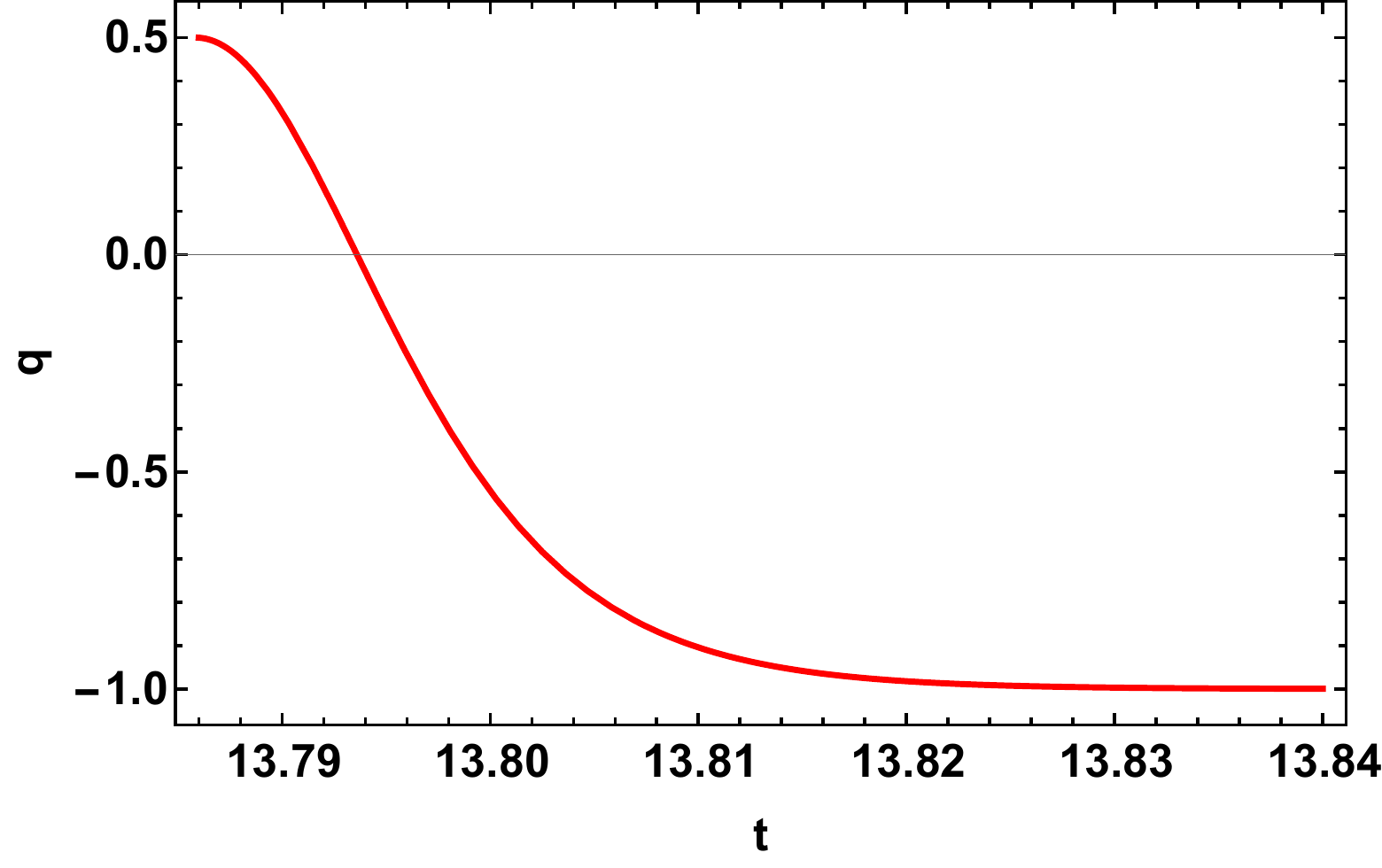}
\caption{Profile of the deceleration parameter vs cosmic time t.}\label{f23}
\end{figure}

In this case, we have obtained constant negative pressure with dark energy EoS parameter follows the $\Lambda$CDM EoS. The effective EoS parameter in the fig.\eqref{f22} converges to the $\Lambda$CDM line and the deceleration parameter in the fig. \eqref{f23} shows a transition from deceleration phase to acceleration phase in the recent past. Hence this cosmological $f(Q)$ model mimics the standard $\Lambda$CDM model.


\section{Conclusion}\label{sec8}

In this article, we attempted to describe the evolution of dark energy from the geometry of spacetime. We considered a $f(Q)$ model which contains a linear and a non-linear form of non-metricity scalar that is $f(Q)=\alpha Q + \beta Q^n$, where $\alpha$, $\beta$, and $n$ are free parameters. Then we derived the expressions for density, deceleration, and the EoS parameters for our cosmological model. We found that for higher positive values of $n$ specifically $n \geq 1$, dark energy fluid part evolving due to non-metricity behaves like quintessence type dark energy while for higher negative values of $n$ specifically $n \leq -1$ it follows phantom scenario. For $n=0$, our cosmological $f(Q)$ model mimics the $\Lambda$CDM model of GR. Furthermore, we have analyzed the behavior of different energy conditions. We obtained that, except SEC all the energy conditions are satisfied for the case $n \geq 1$ while all the energy conditions are violated for the case $n \leq -1$. Therefore, depending upon the choice of $n$ we found that one can obtain quintessence, phantom, and the $\Lambda$CDM like behavior without invoking any dark energy component or exotic fluid in the matter part. Thus, we conclude that the geometrical generalization of GR can be a viable candidate for the description of origin of the dark energy. Also, our work motivates to study reconstruction schemes or to constrain the models of modified theories of gravity via a numerical approach.

\section*{Acknowledgments} \label{sec9}
RS acknowledges University Grants Commission(UGC), New Delhi, India for awarding Junior Research Fellowship (UGC-Ref. No.: 191620096030). AD is supported in part by the FRGS research grant (Grant No. FRGS/1/2021/STG06/UTAR/02/1). We are very much grateful to the honorable referees and to the editor for the
illuminating suggestions that have significantly improved our work in terms
of research quality, and presentation.


\end{document}